\documentclass[%
reprint,
amsmath,amssymb,
aps,
floatfix,
superscriptaddress
]{revtex4-1}

\usepackage{graphicx}					% Include figure files
\usepackage{dcolumn}					% Align table columns on decimal point
\usepackage{bm}						% bold math
\usepackage{float}
\usepackage{gensymb}
\usepackage{multirow}

\renewcommand{\eqref}[1]{Eq.~(\ref{#1})}
\newcommand{\eqrefs}[2]{Eqs.~(\ref{#1}) and (\ref{#2})}

\newcommand{\eqrefsss}[3]{Eqs.~(\ref{#1}), (\ref{#2}), and (\ref{#3})}
\newcommand{\Eqref}[1]{Equation~(\ref{#1})}

\newcommand{\pdf}[2]{\frac{\partial #1}{\partial #2}}

\newcommand{\threevec}[1]{\mathbf{#1}}
\newcommand{\twovec}[1]{\mathbf{#1}}

\renewcommand{\i}{\text{in}}
\renewcommand{\o}{\text{out}}

\newcommand{\B}{\mathrm{f}}
\newcommand{\LB}{g}

\newcommand{\RE}{\text{E}}
\newcommand{\SHE}{\text{sH}}

\newcommand{\avg}[1]{\bar{#1}}

\newcommand{\mhat}{\hat{\threevec{m}}}

\newcommand{\zhat}{\hat{\threevec{z}}}

\newcommand{\khat}{\hat{\threevec{k}}}
\newcommand{\kp}{\twovec{k}_{||}}
\newcommand{\kvec}{\threevec{k}}

\newcommand{\rp}{\twovec{r}_{||}}
\newcommand{\rvec}{\threevec{r}}

\newcommand{\ch}{c}

\newcommand{\dhat}{\hat{\threevec{d}}}
\newcommand{\fhat}{\hat{\threevec{f}}}
\newcommand{\lhat}{\hat{\bm{\ell}}}

\newcommand{\Gm}{G_{\uparrow \downarrow}}

\newcommand{\ImGm}{\text{Im}[\Gm]}

\newcommand{\Ge}{\tilde{G}}
\newcommand{\Gem}{\Ge_{\uparrow \downarrow}}
\newcommand{\ReGem}{\text{Re}[\Gem]}
\newcommand{\ImGem}{\text{Im}[\Gem]}

\newcommand{\magnetization}{\text{mag}}
\newcommand{\lattice}{\text{latt}}
\newcommand{\NM}{\text{NM}}
\newcommand{\FM}{\text{FM}}
\newcommand{\MCT}{\text{MCT}}

\newcommand{\mf}{\text{mf}}
\newcommand{\sfp}{\text{sf}}

\newcommand{\FS}{\text{FS}}

\newcommand{\TDBZ}{\text{2DBZ}}
\newcommand{\TDPC}{\text{2DPC}}

\newcommand{\bulk}{\text{bulk}}
\newcommand{\eq}{\text{eq}}

\newcommand{\lsf}{l_{\text{sf}}}

\newcommand{\EF}{\tilde{E}}

\newcommand{\Rpartial}[1]{\overrightarrow{\partial_#1}}
\newcommand{\Lpartial}[1]{\overleftarrow{\partial_#1}}

\newcommand{\LBE}{f^\RE}

\begin{document}

\preprint{APS/123-QED}

% ------------------------------------------------------------------------------------------------------------------------------------------------------------------------------------------------------------------
% Title
% ------------------------------------------------------------------------------------------------------------------------------------------------------------------------------------------------------------------

\title{Spin Transport at Interfaces with Spin-Orbit Coupling: Formalism}
%\thanks{A footnote to the article title}

% ------------------------------------------------------------------------------------------------------------------------------------------------------------------------------------------------------------------
% Authors and affiliations
% ------------------------------------------------------------------------------------------------------------------------------------------------------------------------------------------------------------------

\author{V. P. Amin}
\email{vivek.amin@nist.gov}
\affiliation{
Maryland NanoCenter, University of Maryland, College Park, MD 20742
}
\affiliation{
Center for Nanoscale Science and Technology, National Institute of Standards and Technology, Gaithersburg, Maryland 20899, USA
}
\author{M. D. Stiles}
\affiliation{
Center for Nanoscale Science and Technology, National Institute of Standards and Technology, Gaithersburg, Maryland 20899, USA
}

%\author{M. D. Stiles}
%\affiliation{
%Center for Nanoscale Science and Technology, National Institute of Standards and Technology, Gaithersburg, Maryland 20899, USA
%}

\date{\today}

% ------------------------------------------------------------------------------------------------------------------------------------------------------------------------------------------------------------------
% Abstract
% ------------------------------------------------------------------------------------------------------------------------------------------------------------------------------------------------------------------

\begin{abstract}

We generalize magnetoelectronic circuit theory to account for spin transfer to and from the atomic lattice via interfacial spin-orbit coupling.  This enables a proper treatment of spin transport at interfaces between a ferromagnet and a heavy-metal non-magnet.  This generalized approach describes spin transport in terms of drops in spin and charge accumulations across the interface (as in the standard approach), but additionally includes the responses from in-plane electric fields and offsets in spin accumulations.  A key finding is that in-plane electric fields give rise to spin accumulations and spin currents that can be polarized in any direction, generalizing the Rashba-Edelstein and spin Hall effects.  The spin accumulations exert torques on the magnetization at the interface when they are misaligned from the magnetization.  The additional out-of-plane spin currents exert torques via the spin-transfer mechanism on the ferromagnetic layer.   To account for these phenomena we also describe spin torques within the generalized circuit theory.  The additional effects included in this generalized circuit theory suggest modifications in the interpretations of experiments involving spin orbit torques, spin pumping, spin memory loss, the Rashba-Edelstein effect, and the spin Hall magnetoresistance.

\end{abstract}

%\begin{description}
%\end{description}

% ------------------------------------------------------------------------------------------------------------------------------------------------------------------------------------------------------------------
% PACS
% ------------------------------------------------------------------------------------------------------------------------------------------------------------------------------------------------------------------

\maketitle

% ------------------------------------------------------------------------------------------------------------------------------------------------------------------------------------------------------------------
% ------------------------------------------------------------------------------------------------------------------------------------------------------------------------------------------------------------------
% Introduction
% ------------------------------------------------------------------------------------------------------------------------------------------------------------------------------------------------------------------
% ------------------------------------------------------------------------------------------------------------------------------------------------------------------------------------------------------------------

\section{Introduction}

The spin-orbit interaction couples the spin and momentum of carriers, leading to a variety of important effects in spintronic devices.  It enables the conversion between charge and spin currents \cite{SHEReviewWunderlich, SHEReviewHankiewicz}, allows for the transfer of angular momentum between populations of spins \cite{SOTTheoryManchon, SOTTheoryManchon2, SOTTheoryMatosAbiague, SOTExpAndo, SOTExpMiron, SPTheoryTserkovnyak, SPExpSanchez}, couples charge transport and thermal transport with magnetization orientation \cite{AMRExpThomson, AMRTheoryMcGuirePotter, TAMRExpGould, ChantisTAMR, AHEReviewNagosa, BauerSpinCaloritronics, TAMSTheoryAmin, TAMSTheoryCarlos}, and results in magnetocrystalline anisotropy \cite{Neel:54,Daalderop:90,Johnson:96}.  Many of these effects already facilitate technological applications.  The development of such applications can be assisted by both predictive (yet complicated) first-principles calculations and clear phenomenological models, which would aid the interpretation of experiments and help to predict device behavior.

In multilayer systems, bulk spin-orbit coupling plays a crucial role in spin transport but the role of interfacial spin-orbit coupling remains largely unknown.  This uncertainty derives from the uncharacterized transfer of angular momentum between carriers and the atomic lattice while scattering from interfaces with spin-orbit coupling.  This transfer of angular momentum occurs because a carrier's spin is coupled via spin-orbit coupling to its orbital moment, which is coupled via the Coulomb interaction to the crystal lattice.  Such interfaces behave as either a sink or a source of spin polarization for carriers in a way that does not yet have an accurate phenomenological description.  In this paper we develop a formal generalization of magnetoelectronic circuit theory to treat interfaces with spin-orbit coupling.  In a companion paper, we extract the most important consequences of this generalization and show that they capture the dominant effects found in more complicated Boltzmann equation calculations.

% ------------------------------------------------------------------------------------------------------------------------------------------------------------------------------
% Figure: Motivation
% ------------------------------------------------------------------------------------------------------------------------------------------------------------------------------
\begin{figure*}[t!]
	\centering
	\vspace{0pt}	
	\includegraphics[width=1\linewidth,trim={0.25cm 1.3cm 0.7cm 0.1cm},clip]{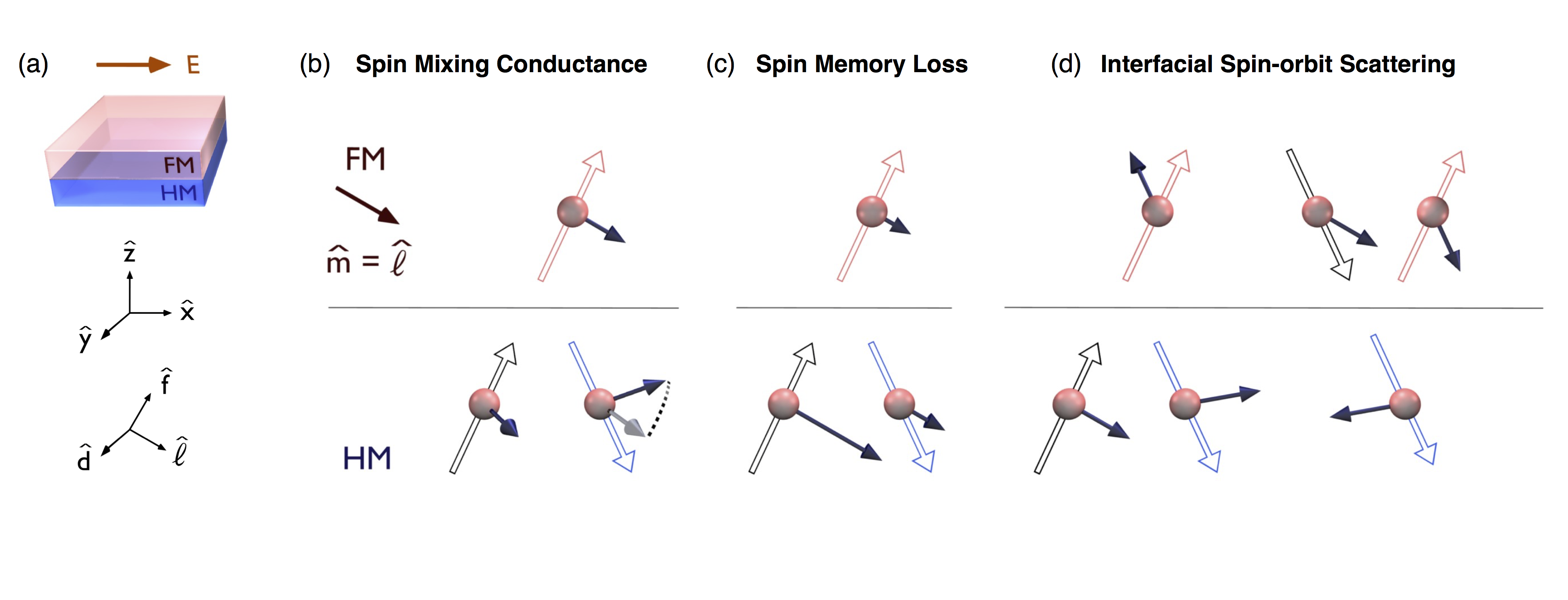}
	\vspace{-12pt}	
	\caption{
	(Color online)  (a)  A heavy metal/ferromagnet bilayer subject to an in-plane electric field.  The axes directly below the bilayer is used to describe electron flow, where the $z$-axis points normal to the interface plane.  The other axes is used to describe spin orientation, where the direction $\ell$ points along the magnetization while the directions $d$ and $f$ span the plane transverse to $\ell$.  (b)  Depiction of the physics described by the spin mixing conductance.  Spins incident from the heavy metal briefly precess around the magnetization when reflecting off of the interface.  The imaginary part of the spin mixing conductance describes the extent of this precession.  Interfacial spin-orbit coupling changes the effective magnetic field seen by carriers during this process in a momentum-dependent way; this alters the precession axis for each carrier and thus modifies the spin mixing conductance.  (c)  Depiction of the loss of spin polarization that carriers experience while crossing interfaces with spin-orbit coupling.  Without interfacial spin-orbit coupling, carriers retain the portion of their spin polarization aligned with the magnetization, but lose the portion polarized transversely to the magnetization due to dephasing processes just within the ferromagnet.  With interfacial spin-orbit coupling, carriers trade angular momentum with the atomic lattice; this leads to changes in all components of the spin polarization.  This phenomenon, known as spin memory loss, affects each component differently.  The panel illustrates only the loss in spin polarization aligned with magnetization.  (d)  Depiction of interfacial spin-orbit scattering in the presence of an in-plane electric field.  Interfacial spin-orbit coupling allows for spins aligned with the magnetization to become misaligned upon reflection and transmission.  For the scattering potential discussed in Sec.~\ref{BPAnalysis}, the spin of a single reflected carrier cancels the spin of a single carrier transmitted from the other side of the interface.  However, a net cancellation of spin is prevented if the total number of incoming carriers differs between sides, as can happen in the presence of in-plane current flow.  This occurs because an in-plane electric field drives two different charge currents within each layer; this forces the number of carriers with a given in-plane momentum to differ on each side of the interface.  The scattered carriers then carry a net spin polarization and a net spin current.  
	}
	\vspace{-8pt}
	\label{fig:motivation}
\end{figure*}
% ------------------------------------------------------------------------------------------------------------------------------------------------------------------------------
% End Figure
% ------------------------------------------------------------------------------------------------------------------------------------------------------------------------------

To understand the impact of interfacial spin-orbit coupling we consider a heavy metal/ferromagnet bilayer, where in-plane currents generate torques on the magnetization through various mechanisms that involve spin-orbit coupling \cite{SOTExpAndo, SOTExpMiron, SOTExpGarello, SOTExpFan, SOTExpAllen, SOTExpEmori}.  For example, bulk spin-orbit coupling converts charge currents in the heavy metal into orthogonally-flowing spin currents, through a process known as the spin Hall effect \cite{SHETheoryDyakonovPerel, SHETheoryHirsch, SHETheoryZhang, SHETheoryMurakami, SHETheorySinova, SHEExpKato, SHEExpWunderlich}.  Upon entering the ferromagnetic layer these spin currents transfer angular momentum to the magnetization through spin transfer torques \cite{STTTheorySlonczewski, STTTheorySlonczewski2, STTTheoryBerger, STTTheoryRalph, STTTheoryStiles}.  Both the spin Hall effect and spin-transfer torques have been extensively studied, but additional sources contribute to the total spin torque.  These remaining contributions arise from interfacial spin-orbit coupling, which enables carriers of the in-plane charge current to develop a net spin polarization at the interface \cite{REETheoryEdelstein, REEExpGanichev, REEExpKato, REEExpSilov, REEExpSanchez}.  In systems with broken inversion symmetry (such as interfaces) the generation of such spin polarization is known as the Rashba-Edelstein effect.  This spin polarization can exert a torque on any magnetization at the interface via the exchange interaction \cite{SOTExpMiron, SOTTheoryHaney}.  A recent experiment suggests that this mechanism can induce magnetization switching alone, without relying on the bulk spin Hall effect \cite{SOTExpEmori}.

The spin torque driven by the Rashba-Edelstein effect is typically studied by confining transport to the two-dimensional interface.  Semiclassical models can capture the direct and inverse Rashba-Edelstein effects \cite{REETheoryGorini, REETheoryGorini2, REETheoryRaimondi, REETheoryShen} in this scenario.  However, such models are not realistic descriptions of bilayers, in which carriers scatter both along and \emph{across} the interface.  Since spin transport across the interface is affected by the transfer of angular momentum to the atomic lattice, the resulting spin torques are modified in ways that two-dimensional models cannot capture.  The various contributions to spin torques in bilayers remain difficult to distinguish experimentally \cite{SOTExpFan, SOTExpAllen} in part because of the lack of models that accurately capture interfacial spin-orbit coupling \cite{SOTTheoryHaney}.

Interfacial spin-orbit coupling may play an important role in other phenomena.  Spin pumping is one example; it describes the process in which a precessing magnetization generates a spin current \cite{SPTheoryTserkovnyak}.  In heavy metal/ferromagnet bilayers, the pumped spin current flows from the ferromagnet into the heavy metal, where the inverse spin Hall effect generates an orthogonal charge current \cite{SPExpMosendz, SPExpNakayama, SPExpFeng, SPExpVlaminck, SPExpZhang2}.  However, because interfacial spin-orbit coupling transfers spin polarization to the atomic lattice, it modifies the pumped spin current as it flows across the interface.  This transfer of spin polarization remains uncharacterized in many systems, thus contributing to inconsistencies in the quantitative interpretation of experiments \cite{SPExpObstbaum, SPExpSanchez, SPExpZhang, SPTheoryLiu}.  Another example, known as the spin Hall magnetoresistance, describes the magnetization-dependent in-plane resistance of heavy metal/ferromagnet bilayers \cite{SMRTheoryChen, SMRExpWeiler, SMRExpHuang, SMRExpNakayama, SMRExpHahn, SMRExpVlietstra, SMRExpAlthammer}.  Currently this effect is attributed to magnetization-dependent scattering at the interface, but may also contain a contribution from interfacial spin-orbit scattering.  The impact of interfacial spin-orbit coupling on these effects remains unclear due to the absence of appropriate models with which to analyze the data.

Magnetoelectronic circuit theory is the most frequently used approach to model spin currents at the interface between a non-magnet and a ferromagnet.  It describes spin transport in terms of four conductance parameters, where drops in spin-dependent electrochemical potentials across the interface play the role of traditional voltages.  However, the theory cannot describe interfaces with spin-orbit coupling because it does not consider spin-flip processes due to spin-orbit coupling at the interface.  Fig.~\ref{fig:motivation}(a) depicts a typical scattering process described by one of these conductance parameters.  Given its success in describing spin transport in normal metal/ferromagnet bilayers, generalizing magnetoelectronic circuit theory to include interfacial spin-orbit coupling would make it a valuable tool for describing heavy metal/ferromagnet bilayers.

%In this approach, spins aligned transverse to the magnetization decouple from those aligned longitudinal to it.  Only spin currents with longitudinal polarization remain conserved when crossing the interface.  However, this decoupling ceases to exist in the presence of interfacial spin-orbit coupling, preventing magnetoelectronic circuit theory from describing these systems in its current form.

%Spin-orbit coupling enables carriers to trade angular momentum with the atomic lattice.  This occurs because a carrier's spin is coupled to its orbital moment (via spin-orbit coupling), which is also coupled to the orbital moments of the atoms that make up the crystal lattice (via the Coulomb interaction).  

To generalize magnetoelectronic circuit theory one must consider all the ways that interfacial spin-orbit coupling potentially affects spin transport.  One such effect, known as spin memory loss, describes a loss of spin current across interfaces due to spin-orbit coupling.  We illustrate a process that contributes to spin memory loss in Fig.~\ref{fig:motivation}(b).  This loss occurs when the atomic lattice at the interface behaves as a sink of angular momentum.  Recent work \cite{SPTheoryChen} incorporates this behavior into a theory for spin pumping, but descriptions of this effect date back to over a decade ago \cite{SMLExpEid, SMLExpKurt, SMLExpNguyen, SMLExpBass}.  Thus generalizing magnetoelectronic circuit theory for interfaces with spin-orbit coupling requires accounting for spin memory loss.  By incorporating spin-flip processes at the interface into magnetoelectronic circuit theory, one can treat this aspect of the phenomenology of interfacial spin-orbit coupling.

Another important consequence of interfacial spin-orbit coupling is that in-plane electric fields can create spin currents that flow away from the interface.  First principles calculations of Pt/Py bilayers suggest that a greatly enhanced spin Hall effect occurs at the interface (as compared to the bulk) that could generate such spin currents \cite{SHETheoryWang}.  This suggests that in-plane electric fields (and not just drops in spin and charge accumulations across the interface) must play a role in generalizations of magnetoelectronic circuit theory.  It also suggests that one cannot confine transport to the two-dimensional interface when describing the effect of in-plane electric fields.  Instead, one must consider transport both along and across the interface.  Some of the consequences of this three-dimensional picture have been investigated in multilayer systems containing an insulator \cite{REETheoryBorge, SSOSTheoryZhang}.  The only semiclassical calculations of three-dimensional metallic bilayers are based on the Boltzmann equation \cite{SOTTheoryHaney}.  Like spin memory loss, these spin currents must be included in generalizations of magnetoelectronic circuit theory to fully capture the effect of interfacial spin-orbit coupling.  In the following we give a semiclassical picture of how such spin currents arise, and how they exert magnetic torques that are typically not considered in bilayers.

Fig.~\ref{fig:motivation}(c) depicts how spins aligned with the magnetization scatter from an interface with spin-orbit coupling.  For the scattering potential discussed in Sec.~\ref{BPAnalysis}, single reflected and transmitted spins cancel on each side of the interface.  However, the \emph{net} cancellation of spin is avoided if the number of incoming carriers differs between sides.  In the simplest scenario, this occurs if the in-plane electric field drives different currents within each layer, so that the occupancy of carriers differs on either side for a given in-plane momentum.  We find that through this mechanism, carriers subject to interfacial spin-orbit scattering can carry a net spin current in addition to exhibiting a net spin polarization.  If the net spin polarization is misaligned with the magnetization, it can exert a torque on the magnetization at the interface.  This describes the contribution to the spin torque normally associated with the Rashba-Edelstein effect (discussed earlier).  However, the spin currents created by interfacial spin-orbit scattering can flow away from the interface, and those that flow into the ferromagnet exert additional torques.  Although these spin currents generate torques via the spin-transfer mechanism, they arise from interfacial spin-orbit scattering instead of the spin Hall effect.  This mechanism, which cannot be captured by confining transport to the two-dimensional interface, is not usually considered when analyzing spin torques in bilayers.  However, it can contribute to the total spin torque in important ways.  For instance, it allows for spin torques generated by interfacial spin-orbit coupling to point in directions typically associated with the spin Hall effect.  The spin polarization and flow directions of these spin currents are not required to be orthogonal to each other or the electric field, unlike the spin currents generated by the spin Hall effect in infinite bulk systems.  More work is needed to determine how this semiclassical description of interfacial spin current generation compares with the first principles description of an enhanced interfacial spin Hall effect \cite{SHETheoryWang}.

In this paper, we generalize magnetoelectronic circuit theory to include interfacial spin-orbit coupling.  Not only does interfacial spin-orbit coupling modify the conductance parameters introduced by magnetoelectronic circuit theory, it requires additional \emph{conductivity} parameters to capture the spin currents that arise from in-plane electric fields and spin-orbit scattering.  Furthermore, the transfer of angular momentum between carriers and the atomic lattice at the interface alters the spin torque that carriers can exert on the magnetization; this introduces additional parameters that are needed to distinguish spin torques from spin currents.  However, we find that many of the parameters in this generalized circuit theory may be neglected when modeling spin-orbit torques in bilayer systems, and that including the conductivity and spin torque parameters is more important than modifying the conductance parameters.  As with magnetoelectronic circuit theory, we provide microscopic expressions for most parameters.  

In a companion paper, to highlight the utility of the proposed theory, we produce an analytical model describing spin-orbit torques caused by the spin-Hall and interfacial Rashba-Edelstein effects.  We achieve this by solving the drift-diffusion equations with this generalization of magnetoelectronic circuit theory.  In that paper, we focus on only the parameters that describe the response of in-plane electric fields, and neglect all other changes to magnetoelectronic circuit theory.  We show that this simplified approach captures the most important effects found in Boltzmann equation calculations of a model system.  In this paper, we discuss the complete generalization of magnetoelectronic circuit theory in the presence of interfacial spin-orbit coupling.

%While much of the recent focus in this field concerns interfaces between ferromagnets and non-magnets, this formalism may be modified to describe interfaces between two non-magnetic materials as well.  One could include disorder by introducing scattering matrix elements that couple neighboring momentum states, also as done in magnetoelectronic circuit theory \cite{Xia:2002}.

In Sec.~\ref{sec:Phenom} of this paper we describe spin transport at interfaces with and without interfacial spin-orbit coupling.  In Sec.~\ref{BPDerivation} we motivate the derivation of all parameters, leaving some details for appendices \ref{ap:SimplifiedForm} and \ref{ap:DerivationTorques}.  In Sec.~\ref{BPAnalysis} we perform a numerical analysis of each boundary parameter for a scattering potential relevant to heavy metal/ferromagnet bilayers.  This analysis allows us to determine which parameters matter the most in these systems.  Finally, in Sec.~\ref{sec:Outlook} we discuss implications of our theory on experiments involving spin orbit torque, spin pumping, the Rashba-Edelstein effect, and the spin Hall magnetoresistance. 

%There we demonstrate that the drift-diffusion approach, coupled with these generalized boundary conditions, gives quantitatively similar results as the Boltzmann approach.  

% ------------------------------------------------------------------------------------------------------------------------------------------------------------------------------------------------------------------
% ------------------------------------------------------------------------------------------------------------------------------------------------------------------------------------------------------------------
% Spin/Charge Transport at Interfaces
% ------------------------------------------------------------------------------------------------------------------------------------------------------------------------------------------------------------------
% ------------------------------------------------------------------------------------------------------------------------------------------------------------------------------------------------------------------

\section{Spin And Charge Transport at Interfaces}
\label{sec:Phenom}

In the following we discuss the general phenomenology of spin transport at interfaces with and without spin orbit coupling.  We first describe some conventional spin transport models to build up to the proposed model, and refrain from presenting explicit expressions of any parameters until later sections.

\subsection{Collinear spin transport}

In the absence of spin-flip processes one often assigns separate current densities for majority ($j_{\uparrow}$) and minority ($j_{\downarrow}$) carriers, i.e.
\begin{align}
\label{eq:ColinearSpinTransportOriginal}
j_{\uparrow} = G_{\uparrow} \Delta \mu_{\uparrow}
\quad \quad \quad
j_{\downarrow} = G_{\downarrow} \Delta \mu_{\downarrow}.
\end{align}
Here $G_{\uparrow / \downarrow}$ denotes the spin-dependent interfacial conductance, while $\Delta \mu_{\uparrow / \downarrow}$ refers to the drop in quasichemical potential for each carrier population across the interface.  We may then define charge ($\ch$) and spin ($s$) components for the drop in quasichemical potential
\begin{align}
\Delta\mu_{\ch} &= \Delta\mu_{\uparrow} + \Delta\mu_{\downarrow}	\\
\Delta\mu_{s} &= \Delta\mu_{\uparrow} - \Delta\mu_{\downarrow},
\end{align}
and for the current densities
\begin{align}
j_{\ch} &= j_{\uparrow} + j_{\downarrow}		\\
j_{s} &= j_{\uparrow} - j_{\downarrow}.
\end{align}
across the interface.  Using the following modified conductance parameters
\begin{align}
G_{\pm} &= \frac{1}{2} \big{(} G_{\uparrow} \pm G_{\downarrow}  \big{)},
\end{align}
we may rewrite \eqref{eq:ColinearSpinTransportOriginal} as
\begin{align}
\label{eq:ColinearSpinTransport}
\begin{pmatrix}
j_s		\\
j_\ch	\\
\end{pmatrix}
=
\begin{pmatrix}
G_+		&		G_-	\\
G_-		&		G_+	\\
\end{pmatrix}
\begin{pmatrix}
\Delta \mu_s \\
\Delta \mu_\ch
\end{pmatrix}
\end{align}
instead.  In this case both spin and charge currents are continuous across the interface.  

\subsection{Magnetoelectronic Circuit Theory}

When describing spin orientation in bulk ferromagnetic systems, the magnetization direction provides a natural spin quantization axes.  However, at the interface between a non-magnet and a ferromagnet, the net spin polarizations of each region need not align.  To account for this, one must consider spins in the non-magnet that point in any direction.  In the ferromagnet, spins are misaligned with the magnetization near the interface but become aligned in the bulk.  This occurs because spins precess incoherently around the magnetization; eventually the net spin polarization transverse to the magnetization vanishes.  In transition metal ferromagnets and their alloys, this dephasing happens over distances smaller than the spin diffusion length.

To describe electron flow and spin orientation in non-magnet/ferromagnet bilayers, we use two separate coordinate systems.  For electron flow, we choose the $x/y$ plane to lie along the interface and the $z$-axis to point perpendicular to it.  The interface is located at the $z$-axis origin, and $z=0^-$ and $z=0^+$ describe the regions just within the non-magnet and ferromagnet respectively.  To describe spin orientation, we choose the direction $\ell$ to be along the magnetization ($\lhat = \mhat$) and the directions $d$ and $f$ to be perpendicular to $\lhat$.  The damping-like ($d$) and field-like ($f$) directions point along the vectors $\dhat \propto \mhat \times [\mhat \times (-\threevec{E} \times \zhat)]$ and $\fhat \propto \mhat \times (-\threevec{E} \times \zhat)$ respectively.  This provides a convenient coordinate system for describing spin-orbit torques, because torques with a damping-like component push the magnetization towards the $-\threevec{E} \times \zhat$ direction, while those with a field-like component force the magnetization to precess about $-\threevec{E} \times \zhat$.

We first define the spin and charge accumulations at the interface ($\mu_\alpha$), where the index $\alpha \in [d,f,\ell,\ch,\ell^*,\ch^*]$ describes the type of accumulation.  The first four indices denote the spin ($d$, $f$, $l$) and charge ($\ch$) accumulations in the non-magnet at $z = 0^-$.  The last two indices describe the spin ($\ell^*$) and charge ($\ch^*$) accumulations in the ferromagnet at $z = 0^+$.  In the ferromagnet we omit spin accumulations aligned transversely to the magnetization, due to the dephasing processes discussed above.  Note that the charge \emph{and} spin components of $\mu_\alpha$ have units of voltage.  We then define the spin and charge current densities flowing out-of-plane ($j_{z\alpha}$) in an identical fashion.  The charge and spin components of $j_{z\alpha}$ have the units of number current density \footnote{To convert these quantities back into their traditional units, one must multiply the components describing charge current densities by $-e$ and those describing spin current densities by $\hbar/2$}.  We refer to $\alpha$ as the spin/charge index.

One may redefine any tensor that contains spin/charge indices in another basis when useful.  For instance, we may write the spin accumulations and spin current densities with longitudinal spin polarization in terms of averages and differences across the interface:
\begin{align}
\Delta \mu_{\ell} &= \frac{1}{2} \big{(} \mu_{\ell} - \mu_{\ell^*} \big{)},		\quad \quad ~
\avg{\mu}_{\ell} = \frac{1}{2} \big{(} \mu_{\ell} + \mu_{\ell^*} \big{)},		\\
\Delta j_{z\ell} &= \frac{1}{2} \big{(} j_{z\ell} - j_{z\ell^*} \big{)},				\quad \quad ~
\avg{j}_{z\ell} = \frac{1}{2} \big{(} j_{z\ell} + j_{z\ell^*} \big{)}.				
\end{align}
We may define similar expressions for the charge accumulations and charge current densities.  As we shall see, this basis ($\alpha \in [d,f,\Delta\ell,\Delta\ch,\avg{\ell},\avg{\ch}]$) provides a more physically transparent representation of all quantities. 

In the absence of interfacial spin-orbit coupling, the spin current polarized along the magnetization direction remains conserved.  However, the spin current with polarization transverse to the magnetization dissipates entirely upon leaving the normal metal.  The interface absorbs part of this spin current, while the remaining portion quickly dissipates within the ferromagnet due to a precession-induced dephasing of spins.  The total loss of spin current then results in a spin transfer torque.  Figure~\ref{fig:GenMCT} depicts this process by use of solutions to the drift-diffusion equations.  In this situation, one may show \cite{MCTBrataas, MCTBrataas2} that the spin and charge current densities at $z = 0^\pm$ become
\begin{align}
\label{eq:IntroMCT}
j_{z\alpha} = G^{\MCT}_{\alpha\beta} \mu_\beta
\end{align}
for a conductance tensor $G^{\MCT}_{\alpha\beta}$ given by
\begin{align}
\label{eq:IntroGMCT}
G^{\MCT} =
\bordermatrix{
~			&	d						&	f						&	\Delta l		&	\Delta c		&	\avg{l}		&	\avg{c}	\cr
d			&	\text{Re}[\Gm]			&	-\text{Im}[\Gm]			&	0			&	0			&	0			&	0		\cr
f			&	\text{Im}[\Gm]			&	\text{Re}[\Gm]			&	0			&	0			&	0			&	0		\cr
\avg{l}		&	0						&	0						&	G_+			&	G_-			&	0			&	0		\cr
\avg{c}		&	0						&	0						&	G_-			&	G_+			&	0			&	0		\cr	
\Delta l		&	0						&	0						&	0			&	0			&	0			&	0		\cr
\Delta c		&	0						&	0						&	0			&	0			&	0			&	0		\cr
}.	
\nonumber \\[0.2cm]
\end{align}
This formalism---known as magnetoelectronic circuit theory---disregards spin currents and accumulations in the ferromagnet with polarization transverse to the magnetization (due to the precession-induced dephasing described above).  This amounts to assuming that the processes occurring in the shaded regions of Fig.~\ref{fig:GenMCT}(b) happen entirely at the interface instead.  While this restriction helps to reduce the number of required parameters, it need not apply to non-ferromagnetic systems or extremely thin ferromagnetic layers.  Note that the rows corresponding to average and discontinuous quantities are switched from the columns corresponding to those quantities.  This is done to emphasize that drops in accumulations cause average currents in magnetoelectronic circuit theory.

\Eqref{eq:IntroGMCT} implies that spin populations polarized transverse to the magnetization decouple from those polarized longitudinal to it.  The charge and longitudinal spin current densities still obey \eqref{eq:ColinearSpinTransport}, whereas the transverse (non-collinear) spin current densities experiences a finite rotation in polarization about the magnetization axis.  Note that the \emph{spin mixing conductance} $\Gm$ governs the latter phenomenon.  In general, one obtains all parameters via integrals of the transmission and/or reflection amplitudes over the relevant Fermi surfaces.

\subsection{Spin transport with interfacial spin orbit coupling}

To generalize magnetoelectronic circuit theory, i.e. \eqref{eq:IntroMCT}, to account  for interfacial spin orbit coupling and in-plane electric fields, we introduce the following expression for the spin and charge current densities at the interface:
\begin{align}
j_{i\alpha} &= G_{i \alpha \beta} \mu_\beta + \sigma_{i \alpha} \EF. 	\label{eq:BC1}
\end{align}
Here we use a scaled electric field defined by $\EF \equiv -E / e$ so that the elements of the tensor $\sigma_{i \alpha}$ have units of conductivity.  Without loss of generality, we assume that the electric field points along the $x$ axis.  

% ------------------------------------------------------------------------------------------------------------------------------------------------------------------------------
% Figure: Motivation Generalize MCT
% ------------------------------------------------------------------------------------------------------------------------------------------------------------------------------
\begin{figure}[t!]
	\centering
	\vspace{0pt}	
	\includegraphics[width=1\linewidth,trim={0.25cm 0.1cm 0.15cm 0.1cm},clip]{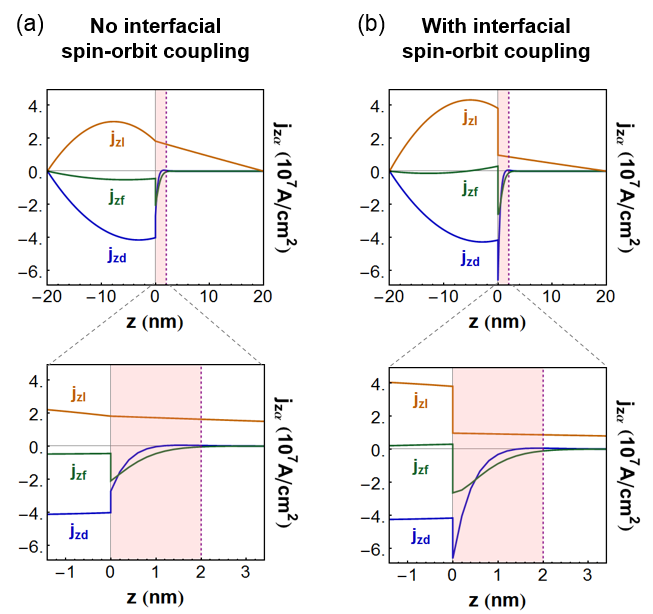}
	\caption{
	(Color online) Spin current densities plotted versus distance from the interface, calculated using the drift-diffusion equations.  Panel (a) treats the case without interfacial spin-orbit coupling using magnetoelectronic circuit theory as boundary conditions, whereas panel (b) treats the case with interfacial spin-orbit coupling by using \eqref{eq:BC1} as boundary conditions instead.  Due to precession-induced dephasing, $j_{zd}$ and $j_{zf}$ dissipate entirely within the ferromagnet some distance from the interface (denoted by the purple dashed line).  With no interfacial spin-orbit coupling, the spin current density polarized along the magnetization ($j_{zl}$) is conserved, while the spin current densities polarized transversely ($j_{zd}$ and $j_{zf}$) exhibit discontinuities at the interface.  With interfacial spin-orbit coupling, all spin currents are discontinuous at the interface.  Furthermore, interfacial spin-orbit coupling introduces additional sources of spin current via the conductivity $\sigma_{i \alpha}$ and torkivity $\gamma^\FM_{\sigma}$ tensors (when an in-plane electric field is present).  These sources may oppose the spin currents that develop in the bulk.  For example, the inclusion of interfacial spin-orbit coupling leads $j_{zf}$ to switch signs near to the interface, as seen by comparing panels (a) and (b).
	}
	\vspace{0pt}
	\label{fig:GenMCT}
\end{figure}
% ------------------------------------------------------------------------------------------------------------------------------------------------------------------------------
% End Figure
% ------------------------------------------------------------------------------------------------------------------------------------------------------------------------------

The explosion of new parameters (relative to magnetoelectronic circuit theory) is an unfortunate consequence of spin-flip scattering at the interface.  Like magnetoelectronic circuit theory, one may express each parameter as an integral of scattering amplitudes over the relevant Fermi surfaces; to discover which parameters may be neglected we numerically study these integrals in Sec.~\ref{BPAnalysis}.  Here, we discuss the overarching implications of this model.  In particular, three new concepts emerge from the above expression:

First of all, the current density $j_{i\alpha}$ now includes an index describing its direction of flow ($i \in [x,y,z]$), which was previously assumed to be out-of-plane.  In this generalization, a buildup of spin and charge accumulation at interfaces may lead to spin and charge currents that flow both in-plane and out-of-plane.  The treatment of in-plane currents close to the interface requires not only the evaluation of \eqref{eq:BC1}, but also an extension of the drift-diffusion equations themselves. 

Secondly, \eqref{eq:BC1} depends on values of the spin and charge accumulations from each side of the interface, rather than differences in those values across the interface.  This suggests that currents result from both drops in accumulations \emph{and} non-zero averages of spin accumulation at the interface \footnote{Note, however, that an average charge accumulation amounts to an offset in electric potential; thus it cannot affect the interfacial spin currents}.

Finally, interfacial spin-orbit scattering results in a conductivity tensor ($\sigma_{i\alpha}$) that drives spin currents in the presence of an in-plane electric field.  This feature represents the greatest conceptual departure from previous theories describing spin transport and is motivated by results from the Boltzmann equation.  Figure~\ref{fig:GenMCT} describes how some of these properties alter solutions of the drift-diffusion equations, as compared with magnetoelectronic circuit theory.  

Without interfacial spin-orbit coupling the in-plane conductance tensors ($G_{x\alpha\beta}$ and $G_{y\alpha\beta}$) vanish, implying that accumulations do not create in-plane currents in this scenario.  The conductivity tensor vanishes as well.  Spin transport transverse to the magnetization still decouples from that longitudinal to it, and magnetoelectronic circuit theory is recovered.  In the presence of interfacial spin-orbit coupling, none of the tensors elements introduced in \eqref{eq:BC1} necessarily vanish, and spin transport in all polarization directions becomes coupled.  However, for the interfacial scattering potential studied in Sec.~\ref{BPAnalysis}, many parameters differ by orders of magnitude; thus certain parameters may be neglected on a situational basis.

\subsection{Spin-orbit torques}

Without interfacial spin-orbit coupling, spin and charge accumulations at an interface create both a spin polarization and spin currents.  The spin polarization develops at $z = 0$ and exerts a torque on any magnetization at the interface via the exchange interaction.  The spin current that develops at $z = 0^+$ exerts an additional torque by transferring angular momentum to the ferromagnetic region via dephasing processes.  For simplicity, we assume that this spin current transfers all of its angular momentum to the magnetization rather than the bulk atomic lattice.  We do so under the assumption that the dephasing processes within the ferromagnet diminish spin currents faster than the spin diffusive processes caused by bulk spin-orbit coupling.  All of the incident transverse spin current is then lost at the interface ($z = 0$) or in the bulk of the ferromagnet ($z > 0$), and carriers can only exchange angular momentum with the magnetization.  Thus the spin current at $z = 0^-$, which represents the incident flux of angular momentum on the magnetized part of the bilayer, equals the total spin torque on the system.  Furthermore, the spin torques at $z = 0$ and $z > 0$ add up to equal the spin current at $z = 0^-$.

However, at interfaces with spin-orbit coupling, the atomic lattice behaves as a reservoir that carriers may transfer angular momentum to.  In this scenario, carriers exert spin torques on both the magnetization and the lattice.  We cannot compute spin torques solely from the spin currents described by \eqref{eq:BC1} if we are to account for the losses to this additional reservoir of angular momentum.  Thus, we introduce a separate expression for the total spin torque on the bilayer:
\begin{align}
\tau_{\sigma} &= \Gamma_{\sigma \beta} \mu_\beta + \gamma_{\sigma} \EF, 		\label{eq:BC2}
\end{align}
Note that the index $\sigma \in [d,f]$ describes the directions transverse to the magnetization, since spin torques only point in those directions.  The tensor $\Gamma$, known as the torkance, describes contributions to the spin torque from the buildup of spin and charge accumulation at an interface.  The tensor $\gamma$, which we call the \emph{torkivity}, captures the corresponding contributions from an external, in-plane electric field.  The torkivity tensor originates from interfacial spin-orbit scattering, much like the conductivity tensor introduced earlier.  

We may separate the total spin torque into two contributions:
\begin{align}
\Gamma_{\sigma \beta}		&=		\Gamma^{\magnetization}_{\sigma \beta}	+	\Gamma^{\FM}_{\sigma \beta}			\label{eq:BC2a}		\\
\gamma_{\sigma}			&=		\gamma^{\magnetization}_{\sigma}		+	\gamma^{\FM}_{\sigma}				\label{eq:BC2b}.		
\end{align}
The first tensors on the right hand side of \eqrefs{eq:BC2a}{eq:BC2b} describe torques exerted by the spin polarization at $z = 0$.  The second tensors describe the spin torque exerted in the bulk of the ferromagnet ($z > 0$).  Both torques are exerted on the magnetization rather than on the atomic lattice.  Here we assume that the torque at $z > 0$ equals the transverse spin current at $z = 0^+$ as before.  Thus, the spin torques exerted at $z = 0$ and $z > 0$ are both included in the torkance and torkivity tensors.

Without interfacial spin-orbit coupling, the torkivity tensor vanishes and the torkance tensor $\Gamma_{\sigma\beta}$ becomes identical to $G_{z\sigma\beta}$.  This indicates that the transverse spin current at $z = 0^-$ equals the total spin torque, as expected.  In the presence of interfacial spin-orbit coupling, the lattice also receives angular momentum from carriers; in this case $\Gamma_{\sigma\beta} \neq G_{z\sigma\beta}$ and $\gamma_{\sigma} \neq 0$.  Thus, by computing the tensors introduced in \eqref{eq:BC2}, one may calculate spin-orbit torques such that the lattice torques are accounted for.  Furthermore, \eqrefs{eq:BC2a}{eq:BC2b} allow one to separate the total spin torque into its interfacial and bulk ferromagnet contributions.

% ------------------------------------------------------------------------------------------------------------------------------------------------------------------------------------------------------------------
% ------------------------------------------------------------------------------------------------------------------------------------------------------------------------------------------------------------------
% Derivation of Boundary Parameters
% ------------------------------------------------------------------------------------------------------------------------------------------------------------------------------------------------------------------
% ------------------------------------------------------------------------------------------------------------------------------------------------------------------------------------------------------------------

\section{Derivation of Boundary Parameters}
\label{BPDerivation}

% ------------------------------------------------------------------------------------------------------------------------------------------------------------------------------
% Figure: K Space Distributions
% ------------------------------------------------------------------------------------------------------------------------------------------------------------------------------
\begin{figure*}[t!]
	\centering
	\vspace{0pt}
	\includegraphics[width=1\textwidth,trim={0.5cm 0.4cm 1.6cm 0.1cm},clip]{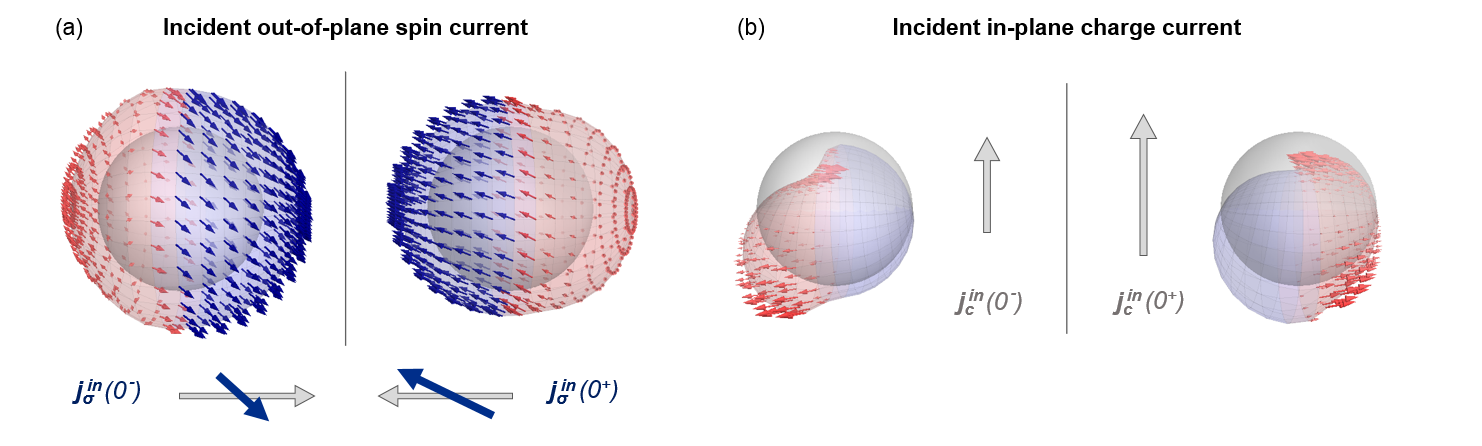}
	\vspace{0pt}
	\caption{
	(Color online) Non-equilibrium distribution functions $\LB_\alpha(\threevec{k})$ in the presence of interfacial spin-orbit scattering, resulting from an (a) incident spin and charge accumulation and an (b) in-plane external electric field.  The images depict $\LB_\alpha(\threevec{k})$ on each side of the interface plotted over $\kvec$-space.  The gray spheres represent the equilibrium Fermi surface.  The colored surfaces represent the non-equilibrium perturbation to the Fermi surface, given by the charge distribution $\LB_\ch(\threevec{k})$  (not to scale).  The arrows denote the spin distribution $\LB_\sigma(\threevec{k})$.  The blue and red regions represent the wavevectors pointing incident and away from the interface respectively.  (a) Scenario in which the incident carriers exhibit a net spin and charge accumulation.  The spin-polarization of the outgoing carriers differs from the incident carriers due to interfacial spin-orbit scattering.  The total spin/charge current density ($j_{i\alpha}$) and the resulting spin torques ($\tau_{\sigma}$) are related to the total spin/charge accumulation ($\mu_\alpha$) by the tensors $G_{i\alpha\beta}$ and $\Gamma_{\sigma\beta}$ respectively.  (b)  Scenario in which the incident carriers are subject to an in-plane electric field.  The in-plane electric field drives two different charge currents on each side of the interface, since each layer possesses a different bulk conductivity.  This shifts the occupancy of carriers (i.e. the charge distribution) differently on each side of the interface.  When spin-\emph{unpolarized} carriers scatter off of an interface with spin-orbit coupling they become spin-polarized.  Because the occupancy of incident carriers was asymmetrically perturbed at the interface, a net cancellation of spin is avoided in even the simplest scattering model.  The resulting spin/charge currents and spin torques are captured by the tensors $\sigma_{i \alpha}$ and $\gamma_{\sigma}$ respectively.  Note that for a ferromagnetic layer, in-plane electric fields also create incident in-plane spin currents as well (suppressed for clarity in this figure).	}
      \vspace{0pt}
	\label{fig:kspace}
\end{figure*}
% ------------------------------------------------------------------------------------------------------------------------------------------------------------------------------
% End Figure
% ------------------------------------------------------------------------------------------------------------------------------------------------------------------------------ 

Interfacial spin-orbit coupling causes both momentum and spin-dependent scattering at interfaces.  If the incident distribution of carriers depends on momentum and/or spin, outgoing carriers may become spin-polarized via interfacial spin-orbit scattering.  This gives rise to non-vanishing accumulations, currents, and torques, which are related by \eqrefs{eq:BC1}{eq:BC2}.  We now motivate these relationships, which can be expressed in terms of scattering amplitudes.  We do so by approximating the non-equilibrium distribution function near the interface.

We first consider the total distribution function $\B_{\alpha} (\threevec{k})$, which gives the momentum-dependent occupancy of carriers described by the spin/charge index $\alpha$.  In equilibrium, this distribution function equals the Fermi-Dirac distribution $\B^{\eq}_\alpha(\varepsilon_{\alpha\threevec{k}})$.  Just out of equilibrium, $\B_{\alpha} (\threevec{k})$ is perturbed as follows
\begin{align}
\B_{\alpha} (\threevec{k}) = \B^{\eq}_\alpha(\varepsilon_{\alpha\threevec{k}}) + \pdf{\B^{\eq}_\alpha}{\varepsilon_{\alpha\threevec{k}}} \LB_{\alpha}(\threevec{k}),
\label{eq:BoltDist}
\end{align}
where $\LB_{\alpha}(\threevec{k})$ denotes the non-equilibrium distribution function.  The equilibrium distribution functions vanish for $\alpha \in [d,f,\ell]$ since the non-magnet exhibits no equilibrium spin polarization.  However, the non-equilibrium distribution functions for all spin/charge indices are generally non-zero.  

To obtain the tensors introduced in \eqrefs{eq:BC1}{eq:BC2}, we must evaluate $\LB_{\alpha}(\threevec{k})$ near the interface.  One could evaluate $\LB_{\alpha}(\threevec{k})$ by solving the spin-dependent Boltzmann equation for the bilayer system.  In this approach one captures spin transport both in the bulk and at the interface.  A more practical approach is to assume some generic form for $\LB_{\alpha}$ near the interface that is physically plausible.  This approach yields boundary conditions suitable for simpler bulk models of spin transport such as the drift-diffusion equations.  In the companion paper, we show that solving the drift-diffusion equations using these boundary conditions produces quantitatively similar results to solving the Boltzmann equation.

For simplicity, we assume that spherical Fermi surfaces describe carriers in both layers.  Later we generalize this formalism to describe non-trivial electronic structures.  In the non-magnet, all carriers belong to the same Fermi surface.  In the ferromagnet, majority ($\uparrow$) and minority ($\downarrow$) carriers belong to different Fermi surfaces.  Thus we use the spin/charge basis $\alpha \in [d,f,\ell,\ch,\uparrow,\downarrow]$, since in this model carriers belonging to those populations have well-defined Fermi surfaces and velocities.  The tensors derived in this section may be expressed in other spin/charge bases by straightforward linear transformations.

%To obtain the non-equilibrium distribution function $\LB_{\alpha}$ one typically solves the spin-dependent Boltzmann equation.  However, to obtain standard expressions for the tensors in \eqrefs{eq:BC1}{eq:BC2}, we must assume some generic form for $\LB_{\alpha}$ that is physically plausible:
To approximate $\LB_{\alpha}$ at the interface we use the following expression:
\begin{align}
\label{eq:IncDist}
\LB^{\i}_{\alpha}(\kp) = -e \Big{(} q_\alpha + \EF \LBE_{\alpha}(\kp) \Big{)},
\end{align}
\Eqref{eq:IncDist} represents the portion of $\LB_{\alpha}$ incident on the interface, where $\kp$ denotes the in-plane momentum vector and $e$ equals the elementary charge.  The right hand side of \eqref{eq:IncDist} describes two pieces of the incoming distribution function; Fig.~\ref{fig:kspace} depicts both pieces over $\kvec$-space for each side of the interface.  The first term captures spin/charge currents incident on the interface.  They may arise, for example, from the bulk spin Hall effect or ferromagnetic leads.  The quantities $q_\alpha$ denote the isotropic spin/charge polarization of those currents.  The second term represents the anisotropic contribution to the distribution function caused by an external electric field.  We remind the reader that the scaled electric field $\EF$ points along the $x$ axis.  The simplest approximation for $\LBE_{\alpha}(\kp)$ is to use the particular solution of the Boltzmann equation in the relaxation time approximation:
\begin{equation}
\LBE_{\alpha}(\kp) = -e v_{x \alpha}(\kp) \times
\begin{cases} 
0					 		\quad \quad~\;		\alpha \in [d,f,\ell]				\\[0.15cm]
\tau							\quad \quad~\;		\alpha = \ch					\\[0.15cm]
\tau^\uparrow 				\quad \quad\,			\alpha = \; \uparrow			\\[0.15cm]
\tau^\downarrow	 			\quad \quad\,			\alpha = \; \downarrow			
\end{cases}
\label{eq:apfSimple}
\end{equation}
This term describes the in-plane charge current caused by the external electric field, but also describes an in-plane spin current polarized opposite to the magnetization in the ferromagnet.  The momentum relaxation times in the ferromagnet differ between majority ($\tau^\uparrow$) and minority ($\tau^\downarrow$) carriers.  In the non-magnet, the momentum-relaxation time ($\tau$) is renormalized by bulk spin-flip processes (see appendix \ref{ap:SimplifiedForm}).  

The outgoing distribution function
\begin{align}
\label{eq:OutDist}
%\LB^{\o}_{\alpha}(\kp) = \frac{|v_{z\alpha}(\kp)|}{|v_{z\beta}(\kp)|} S'_{\alpha \beta}(\kp) \LB^{\i}_{\beta}(\kp),
\LB^{\o}_{\alpha}(\kp) = S_{\alpha \beta}(\kp) \LB^{\i}_{\beta}(\kp),
\end{align}
is specified by the incoming distribution function and the unitary scattering coefficients $S_{\alpha \beta}$, given by
\begin{align}
S_{\alpha \beta} \equiv \frac{|v_{z\alpha}(\kp)|}{|v_{z\beta}(\kp)|} S'_{\alpha \beta}(\kp),
\label{eq:SIntUnitary}
\end{align}
where
\begin{align}
S'_{\alpha \beta} =
\begin{cases} 
\frac{1}{2} \mathrm{tr} \big{[} r^\dagger \sigma_\alpha r \sigma_\beta \big{]}					~~\quad \quad \quad \: \!		\alpha, \beta \in [d,f,\ell,\ch]										\\[0.3cm]
\frac{1}{2} \mathrm{tr} \big{[} t^\dagger \sigma_\alpha t \sigma_\beta \big{]}					~~\quad \quad \quad \, \,		\alpha \in [d,f,\ell,\ch], ~~ \beta \in [\uparrow,\downarrow]			\\[0.3cm]
\frac{1}{2} \mathrm{tr} \big{[} (t^*)^\dagger \sigma_\alpha t^* \sigma_\beta \big{]}			~~\quad \; \, \,				\alpha \in [\uparrow,\downarrow], ~~ \beta \in [d,f,\ell,\ch]			\\[0.3cm]
\frac{1}{2} \mathrm{tr} \big{[} (r^*)^\dagger \sigma_\alpha r^* \sigma_\beta \big{]}			~~\quad \; \,					\alpha, \beta \in [\uparrow,\downarrow]							
\end{cases} \nonumber \\
\label{eq:SIntPrime}
\end{align}
Here we define the Pauli vector $\sigma_\alpha$ such that $\sigma_d = \sigma_x$, $\sigma_f = \sigma_y$, $\sigma_\ell = \sigma_z$, and
\begin{align}
\sigma_\ch =
\begin{pmatrix}
1		&		0		\\
0		&		1				
\end{pmatrix},	
\quad
\sigma_\uparrow =
\begin{pmatrix}
1		&		0		\\
0		&		0				
\end{pmatrix},	
\quad
\sigma_\downarrow =
\begin{pmatrix}
0		&		0		\\
0		&		1				
\end{pmatrix}.
\end{align}
The coefficient $S'_{\alpha\beta}$ gives the strength of scattering for carriers with spin/charge index $\beta$ into those with spin/charge index $\alpha$.  The scattering coefficients depend on the $2 \times 2$ reflection and transmission matrices for spins pointing along the magnetization axis.  In particular, the matrices $r^*$ and $t^*$ describe reflection and transmission respectively into the ferromagnet.  The matrices $r$ and $t$ describe reflection and transmission into the non-magnet.  Note that the density of states and Fermi surface area element differ between incoming and outgoing carriers.  Thus to conserve particle number one must include the ratio of velocities within the scattering coefficients, as done in \eqref{eq:SIntUnitary}.

We obtain all non-equilibrium quantities near the interface by integrating $\LB_\alpha$ over the relevant Fermi surfaces.  We note that the outgoing part of $\LB_\alpha$ includes the consequences of interfacial scattering, since it depends on the scattering coefficients.  For example, the interfacial exchange interaction leads to spin-dependent scattering, which is captured by the difference in the diagonal elements of the $2 \times 2$ reflection and transmission matrices.  On the other hand, the interfacial spin-orbit interaction introduces spin-flip scattering, which is captured by the off-diagonal elements within these matrices.  Thus, to describe the consequences of interfacial spin-orbit scattering we must not limit the form of the reflection and transmission matrices as was often done in the past. 

We write the current density $j_{i\alpha}$ for carriers with spin/charge index $\alpha$ flowing in direction $i \in [x,y,z]$ as follows:
\begin{align}
j_{i\alpha} &=  \frac{1}{\hbar (2\pi)^3} \frac{1}{v_{F\alpha}} \int_{\FS_\alpha} d^2k v_{i \alpha}(\kvec) \LB_{\alpha}(\kvec) 				\label{eq:DefCurrent}		
\end{align}
Note that all integrals run over the Fermi surface corresponding to the population with spin/charge index $\alpha$.  The quantity $v_{F\alpha}$ denotes the Fermi velocity for that population.  To define the accumulations $\mu_{\alpha}$ we follow the example of magnetoelectronic circuit theory \cite{MCTBrataas, MCTBrataas2} and assume that the incoming currents behave as if they originate from spin-dependent reservoirs.  This implies that the incoming polarization $q_\alpha$ approximately equals the accumulation $\mu_\alpha$ at the interface.  

We have now discussed the requirements for deriving the conductance and conductivity tensors found in \eqref{eq:BC1}.  We obtain these tensors by plugging \eqrefs{eq:IncDist}{eq:OutDist} into \eqref{eq:DefCurrent} and noting that $q_\alpha \approx \mu_\alpha$.  In doing so we write the currents $j_{i\alpha}$ in terms of the accumulations $\mu_\alpha$ and the in-plane electric field $\EF$.  From the resulting expressions one then obtains formulas for the conductance and conductivity tensors in terms of the interfacial scattering coefficients.  We outline this remaining process in appendix \ref{ap:SimplifiedForm}.  In appendix \ref{ap:GeneralForm} we generalize those expressions for the case of non-trivial electronic structures, which allows one to compute the conductance and conductivity tensors for realistic systems.

Having discussed the currents that arise from interfacial spin-orbit scattering, we now discuss the spin torques caused by the same phenomenon.  The transverse spin polarization at $z = 0$ exerts a torque on any magnetization at the interface via the exchange interaction.  The transverse spin current at $z = 0^+$ exerts a torque by transferring angular momentum to the ferromagnet.  The total spin torque then equals the sum of these two torques.  To describe the spin torque at $z = 0$, we must compute the spin polarization at the interface.  To accomplish this we define the following matrix
\begin{equation}
T_{\sigma \beta} =
\begin{cases} 		
\frac{1}{2} \mathrm{tr} \big{[} (t^*)^\dagger \sigma_\sigma t^* \sigma_\beta \big{]}			\quad \quad \: \!				\beta \in [d,f,\ell,\ch]						\\[0.3cm]
\frac{1}{2} \mathrm{tr} \big{[} t^\dagger \sigma_\sigma t \sigma_\beta \big{]}					\quad \quad \quad ~ \;\,		\beta \in [\uparrow,\downarrow]				
\end{cases}
\label{eq:SPerp}
\end{equation}
which describes phase-coherent transmission from all populations into transverse spin states at the interface.  We may then compute the ensemble average of spin density $\langle s_\sigma \rangle$ at $z = 0$ as follows:
\begin{align}
\langle s_\sigma \rangle &= \frac{1}{\hbar (2\pi)^3} \sum_\beta \frac{1}{v_{F\beta}} \int_{\FS_\beta \in \i} d^2k T_{\sigma \beta}(\kp) \LB^{\i}_{\beta}(\kp).			
\nonumber	\\	
\label{eq:DefTorque}
\end{align}
The torque at $z = 0$ is then given by 
\begin{align}
\tau^\magnetization_\sigma =  - \int\limits_{0^-}^{0^+} dz \frac{J_\text{ex}}{\hbar} \big{[} \langle \threevec{s} \rangle \times \mhat \big{]}_\sigma,			\label{eq:DefTorque2}
\end{align}
where $J_\text{ex}$ equals the exchange energy at the interface.  We evaluate this integral over the region that describes the interface, where the exchange interaction and strong spin-orbit coupling overlap.  Note that the cross product $\big{[} \langle \threevec{s} \rangle \times \mhat \big{]}_\sigma = \epsilon_{\sigma\sigma'} \langle s_{\sigma'} \rangle$ is evaluated by computing \eqref{eq:DefTorque}.  

To describe the spin torque at $z = 0^+$, we introduce an additional scattering matrix:
\begin{equation}
\bar{S}_{\sigma \beta} =
\begin{cases} 
\frac{1}{2} \mathrm{tr} \big{[} (t^*)^\dagger \sigma_\sigma t^* \sigma_\beta \big{]}			\quad \quad \: 		\beta \in [d,f,\ell,\ch]				\\[0.3cm]
\frac{1}{2} \mathrm{tr} \big{[} (r^*)^\dagger \sigma_\sigma r^* \sigma_\beta \big{]}			\quad \quad			\beta \in [\uparrow,\downarrow]			
\end{cases}
\label{eq:SIntTransverse}
\end{equation}
This scattering matrix is used to calculate the transverse spin current at $z = 0^+$.  Since this spin current rapidly dephases, it contributes entirely to the spin torque exerted on the ferromagnet.  Note that the currents discussed previously corresponded to carriers with well-defined velocities.  However, transverse spin states in the ferromagnet consist of linear combinations of majority and minority spin states.  Since these spin states possess different phase velocities, the velocities of transverse spin states oscillate over position.  These states also posses different group velocities, and wave packets with transverse spin travel with the average group velocity. The transverse spin current at $z = 0^+$ then equals
\begin{align}
\tau^\FM_\sigma &=  \frac{1}{\hbar (2\pi)^3} \sum_\beta \int_{\TDBZ} d\kp \frac{\avg{v}_{z}(\kp)}{v_{z\beta}(\kp)} \bar{S}_{\sigma \beta}(\kp) \LB^{\i}_{\beta}(\kp),		
\nonumber \\		
\label{eq:DefTorqueFM}
\end{align}
where
\begin{align}
\avg{v}_{z}(\kp)		\equiv		\frac{1}{2}		\Big{(}	v_{z\uparrow}(\kp)		+		v_{z\downarrow}(\kp)	\Big{)}			\label{eq:TransverseVelocityFM}
\end{align}
gives the average group velocity of carriers in the ferromagnet.  Note that we write this integral over the maximal two-dimensional Brillouin zone common to all carriers (see appendices \ref{ap:SimplifiedForm} and \ref{ap:DerivationTorques}).  The total torque then equals the sum of torques at the interface and in the bulk ferromagnet:
\begin{align}
\tau_\sigma = \tau^\magnetization_\sigma + \tau^\FM_\sigma.
\label{eq:TotalTorqueDef}
\end{align}

As before we assume that the incoming polarizations approximately equal the accumulations at the interface.  Thus we obtain $\tau^\magnetization_{\sigma}$ and $\tau^\FM_{\sigma}$ in terms of $\mu_\alpha$ and $\EF$ by plugging \eqrefs{eq:IncDist}{eq:OutDist} into \eqrefsss{eq:DefTorque}{eq:DefTorque2}{eq:DefTorqueFM}.  From the resulting expressions we may define the torkance and torkivity tensors introduced in \eqref{eq:BC2}.  In appendix \ref{ap:DerivationTorques} we discuss this process, and in appendix \ref{ap:GeneralForm} we present generalized expressions for non-trivial electronic structures.

We note that the conductance and conductivity tensors describe the charge current and longitudinal spin current in the ferromagnet, but not the transverse spin currents.  In the ferromagnet, the transverse spin currents dissipate not far from the interface, while the charge current and longitudinal spin current can propagate across the entire layer.  Thus the transverse spin currents in the ferromagnet are best described as spin torques given by $\tau^\FM_\sigma$; this explains why we include them in the torkance and torkivity tensors instead of the conductance and conductivity tensors.  If we derive a similar formalism to describe a non-magnetic bilayer, spin currents polarized in all directions should be included in the conductance and conductivity tensors.  With no magnetism, no spin torques are exerted at or near the interface and the torkance and torkivity tensors are not meaningful.

%\Eqrefs{eq:IncDist}{eq:OutDist} approximate the non-equilibrium distribution function at an interface subject to spin-orbit scattering.  \Eqref{eq:IncDist} assumes that the spin-dependent chemical potential and external electric field entirely determine the incoming distribution function; the interfacial scattering coefficients then determine the outgoing distribution function.  However, the resulting distribution function does not satisfy the Boltzmann equation with the proper boundary conditions at other interfaces (the outer interfaces for example).   Thus we may obtain more accurate boundary parameters by allowing the bulk solutions and interfacial scattering to alter the \emph{incoming} distribution function as well.  The most significant improvement comes from altering the second term on the right hand side of \eqref{eq:IncDist}, which describes the dependence on external electric field alone.  In the companion paper we describe how to calculate these contributions, which enable further agreement between the drift-diffusion and Boltzmann approaches.

% ------------------------------------------------------------------------------------------------------------------------------------------------------------------------------------------------------------------
% Boundary Parameter Analysis
% ------------------------------------------------------------------------------------------------------------------------------------------------------------------------------------------------------------------

\section{Numerical Analysis of Boundary Parameters}
\label{BPAnalysis}

% ------------------------------------------------------------------------------------------------------------------------------------------------------------------------------
% Figure: Boundary Parameters
% ------------------------------------------------------------------------------------------------------------------------------------------------------------------------------
\begin{figure*}[t!]
	\centering
	\includegraphics[width=1\textwidth,trim={0cm 0cm -0.75cm 0cm},clip]{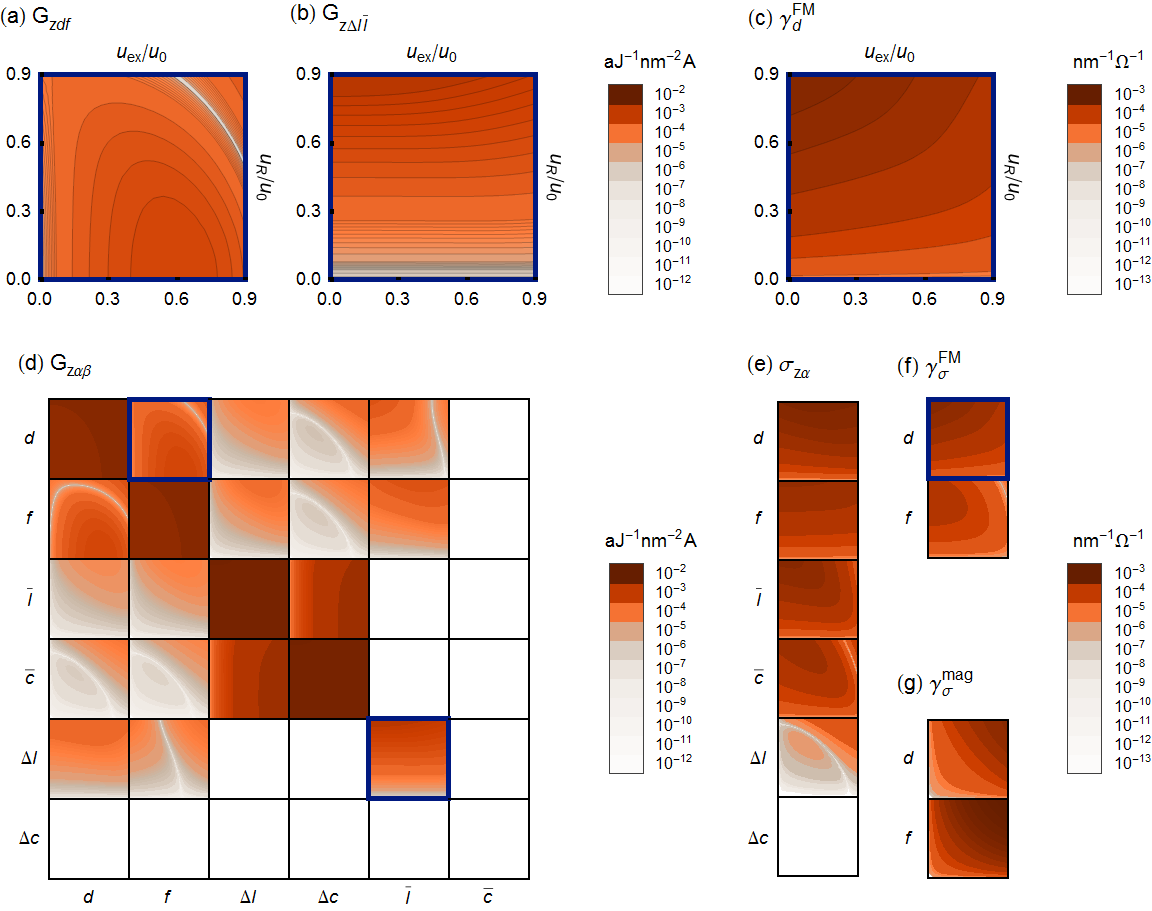}
      \vspace{-2pt}
	\caption{
	(Color online) Contour plots of various boundary parameters versus the interfacial exchange ($u_\text{ex}$) and Rashba ($u_R$) strengths.  The magnetization points away from the electric field $45^o$ in-plane and $22.5^o$ out-of-plane.  Note that the parameters plotted in panels (a)-(c) describe the scattering processes illustrated in Figs.~\ref{fig:motivation}(a)-(c).  (a) Plot of $G_{zdf}$, which generalizes $\ImGm$ in the presence of interfacial spin-orbit coupling.  It describes a rotation of spin currents polarized transversely to the magnetization.  (b) Plot of $G_{z\bar{l}\Delta l}$, which contributes to spin memory loss longitudinal to the magnetization.  It varies mostly with $u_R$, since interfacial spin-orbit coupling provides a sink for angular momentum.  (c) Plot of $\gamma^\FM_{d}$, which describes the out-of-plane, damping-like spin current created by an in-plane electric field and spin-orbit scattering.  It exceeds its field-like counterpart ($\gamma^\FM_{f}$); thus, the resulting spin current exerts a (mostly) damping-like spin torque upon entering the ferromagnet.  (d) An array of contour plots, with each plot shown over an identical range as those in (a)-(c).  The plot in row $\alpha$ and column $\beta$ corresponds to the parameter $G_{z\alpha\beta}$.  From this one may visualize the coupling between spin/charge indices for this tensor, shown across the parameter space of the scattering potential given by \eqref{eq:ScatPot}.  The overall structure of $G_{z\alpha\beta}$ resembles that of magnetoelectronic circuit theory, given by \eqref{eq:IntroGMCT}.  The corresponding figures for (e) $\sigma_{z\alpha}$, (f) $\gamma^\FM_\sigma$, and (g) $\gamma^\magnetization_\sigma$ are also shown.
	}
	\label{fig:G_Minus_Plus}
      \vspace{-2pt}
\end{figure*}
% ------------------------------------------------------------------------------------------------------------------------------------------------------------------------------
% End Figure
% ------------------------------------------------------------------------------------------------------------------------------------------------------------------------------

In the following we numerically analyze the boundary parameters introduced in \eqrefs{eq:BC1}{eq:BC2} in the presence of an interfacial exchange interaction and spin-orbit scattering.  We do so to provide intuition as to the relative strengths of each boundary parameter.  We use a scattering potential localized at the interface \cite{SOTTheoryHaney} that is based on the Rashba model of spin orbit coupling
\begin{align}
\label{eq:ScatPot}
V(\threevec{r}) = \frac{\hbar^2 k_F}{m} \delta(z) \big{(} u_0 + u_\text{ex} \threevec{\sigma} \cdot \mhat + u_R \threevec{\sigma} \cdot (\khat \times \zhat) \big{)}
\end{align}
where $u_0$ represents a spin-independent barrier, $u_\text{ex}$ governs the interfacial exchange interaction, and $u_R$ denotes the Rashba interaction strength.  Plane waves comprise the scattering wavefunctions in both regions.

In Fig.~\ref{fig:G_Minus_Plus} we plot various boundary parameters versus the exchange interaction strength ($u_\text{ex}$) and the Rashba interaction strength ($u_R$).  Figures~\ref{fig:G_Minus_Plus}(a)-(c) display individual boundary parameters, while Figs.~\ref{fig:G_Minus_Plus}(d)-(g) display multiple boundary parameters for a given tensor.  The plots in Figs.~\ref{fig:G_Minus_Plus}(d)-(g) are arranged as arrays to help visualize the coupling between spin/charge components.  The spin-orbit interaction misaligns the preferred direction of spins from the magnetization axis.  Thus, no two tensor elements are identical, though many remain similar.  As expected, the coupling between the transverse spin components and the charge and longitudinal spin components does not vanish.  

The conductance tensor $G_{z\alpha\beta}$ generalizes $G^{\MCT}_{\alpha\beta}$ in the presence of interfacial spin-orbit coupling.  Comparison to \eqref{eq:IntroGMCT} suggests that the parameters $G_{zdd}$ and $G_{zdf}$ represent the real and imaginary parts of a generalized mixing conductance ($\Gem$).  Each element of the conductance tensor experiences a similar perturbation due to spin-orbit coupling.  However, the tensor elements from the $2 \times 2$ off-diagonal blocks in Fig.~\ref{fig:G_Minus_Plus}(d) either vanish or remain two orders of magnitude smaller than those from the diagonal blocks.  This remains true even for values of $u_R$ approaching the spin-independent barrier strength $u_0$.  While these blocks are small for the simple model treated here, they may become important for particular realistic electronic structures.  The fact that the elements $G_{z \Delta\ch \alpha}$ and $G_{z \alpha \avg{\ch}}$ vanish for all $\alpha$ ensures the conservation of charge current and guarantees no dependence on an offset to the charge accumulations.  Note that four additional parameters vanish in the conductance tensor shown in Fig.~\ref{fig:G_Minus_Plus}(d); this occurs because identical scattering wavefunctions were used for both sides of the interface when computing the scattering coefficients.  These parameters do not vanish in general.

The results shown in Fig.~\ref{fig:G_Minus_Plus} were computed for a magnetization with out-of-plane components.  In magnetoelectronic circuit theory, the parameters are independent of the magnetization direction.  With interfacial spin-orbit coupling, this is no longer the case.  In general all of the parameters in \eqrefs{eq:BC1}{eq:BC2} depend on the magnetization direction.  However, we find that this dependence is weak for the model we consider here.   For in-plane magnetizations (not shown) the $2 \times 2$ off-diagonal blocks vanish, but spin-orbit coupling still modifies the diagonal blocks in the manner described above.

In the presence of interfacial spin-orbit coupling the lattice also receives angular momentum from carriers.  This results in a loss of spin current across the interface, or spin memory loss, which the elements $G_{z\Delta l \alpha}$ partly characterize.  The computation of these parameters for realistic electronic structures should help predict spin memory loss in experimentally-relevant bilayers.  In particular, spin memory loss might play a crucial role when measuring the spin Hall angle of heavy metals via spin-pumping from an adjacent ferromagnet \cite{SPExpSanchez}.  Here $G_{z \Delta l \avg{l}}$ provides the strongest contribution to spin memory loss that is caused by accumulations, and approaches the imaginary part of the generalized mixing conductance in magnitude.

Until now, we have discussed the tensors that describe how accumulations affect transport.  However, in-plane electric fields and spin-orbit scattering create additional currents that form near the interface.  In particular, the conductivity parameters $\sigma_{i\alpha}$ describe the currents that can propagate into either layer without significant dephasing.  For instance, the element $\sigma_{z\avg{l}}$ describes an out-of-plane longitudinal spin current driven by an in-plane electric field.  The element $\sigma_{z\Delta l}$ then gives the discontinuity in this spin current across the interface.  This discontinuity arises because of coupling to the lattice, and thus contributes to spin memory loss.

Likewise, the torkivity tensors describe contributions to the total spin torque that arise from in-plane electric fields and spin-orbit scattering.  This includes the torques exerted by the spin polarization at $z = 0$ and by the transverse spin currents at $z = 0^+$.  The tensors $\gamma^\magnetization_{\sigma}$ and $\gamma^\FM_{\sigma}$ describe these torques respectively.  Since the transverse spin currents at $z = 0^+$ quickly dephase in the ferromagnet, we treat them as spin torques and do not include them in the conductivity tensor.  

To understand how the boundary parameters contribute to spin-orbit torques, we note that $\gamma^\magnetization_f > \gamma^\magnetization_d$ over the swept parameter space.  This implies that the torque exerted at $z = 0$ is primarily field-like, which agrees with previous studies of interfacial Rashba spin orbit torques \cite{SOTTheoryHaney}.  However, we also find that $\gamma^\FM_d > \gamma^\FM_f$ for strong $u_\text{R}$; in this case the resulting spin current exerts a damping-like torque by flowing into the ferromagnet.  Both spin torque contributions result from the interfacial Rashba interaction.  This implies that interfacial spin-orbit scattering provides a crucial mechanism to the creation of damping-like Rashba spin torques.  In the companion paper we support this claim by comparing spin-orbit torques computed using both the drift-diffusion and Boltzmann equations.

\section{Outlook}
\label{sec:Outlook}

In the previous section we demonstrated that only certain boundary parameters remain important when modeling spin orbit torques.  The interfacial conductivity and torkivity parameters capture physics due to in-plane external electric fields.  They depend on the difference in bulk conductivities, which are typically easier to estimate than interfacial spin/charge accumulations.  For this reason, calculating the conductivity and torkivity tensors for a realistically-modeled system should provide direct insight into its spin transport behavior.  In particular, we showed that conductivity and torkivity parameters strongly indicate the potential to produce damping-like and field-like torques.  Further studies may yield significant insight into the underlying causes of these and other phenomena for specific material systems.  Even so, treating the elements of these tensors as phenomenological parameters should benefit the analysis of a variety of experiments, which we discuss now.

\emph{(1) Spin pumping/memory loss} --- Spin pumping experiments in Pt-based multilayers suggest that the measured interfacial spin current differs from the actual spin current in Pt, leading to inconsistent predictions of the spin Hall angle \cite{SPExpSanchez,SPExpZhang}.  Rojas-Sanchez et al. \cite{SPExpSanchez} explain this discrepancy in terms of spin memory loss while Zhang et al. \cite{SPExpZhang} attribute it to interface transparency.  The latter characterizes the actual spin current generated at an interface when backscattering is accounted for; it depends on $\Gm$ and does not require interfacial spin-orbit coupling.  Though further experimental evidence is needed to resolve these claims, the elements of $G_{z\alpha\beta}$ characterize both spin memory loss and transparency.  Figure~\ref{fig:G_Minus_Plus} implies that transparency depends on interfacial spin-orbit coupling, while spin memory loss also depends on the interfacial exchange interaction.  Thus, the generalized boundary conditions introduced here unify these two interpretations and allow for further investigation using a single theory.

\emph{(2) Rashba-Edelstein effect} --- Sanchez et al.~\cite{REEExpSanchez} measure the inverse Rashba-Edelstein effect in an Ag/Bi interface, in which interfacial spin orbit coupling converts a pumped spin current into a charge current.  The theoretical methods that describe this phenomena to date \cite{REETheoryEdelstein, REETheoryGorini2, REETheoryRaimondi, REETheoryBorge, REETheoryShen} assume orthogonality between the directional and spin components of the spin current tensor.  However, the conductivity tensor introduced here is robust in general; this implies that interfacial spin-orbit scattering converts charge currents into spin currents with polarization and flow directions not orthogonal to the charge current.  Onsager reciprocity implies that spin currents should give rise to charge currents at the interface that flow in all directions as well.  Thus, the conductivity tensor describes a generalization of the direct and inverse Rashba-Edelstein effects as they pertain to interfaces with spin-orbit coupling.  

\emph{(3) Spin Hall magnetoresistance} --- The conductivity tensor also leads to in-plane charge currents.  These currents depend on magnetization direction via the scattering amplitudes, and thus suggest a new contribution to the spin Hall magnetoresistance based on the Rashba effect in addition to that from the spin Hall effect.  Preliminary calculations of this mechanism suggest a magnetoresistance in Pt/Co of a few percent, which is comparable or greater than experimentally measured values in various systems \cite{SMRExpNakayama, SMRExpHahn, SMRExpVlietstra, SMRExpAlthammer}.

We expect that the most useful approach for interpreting experiments as above is to treat the new transport parameters as fitting parameters.  In the future, this approach can be checked by calculating the parameters from first principles \cite{IntResStiles,Xia:2002} as has been done for magnetoelectronic circuit theory.  This requires computing the boundary parameters for realistic systems using the expressions given in appendix \ref{ap:GeneralForm}.  Such calculations would provide a useful bridge between direct first-principles calculations of spin torques \cite{SOTTTheoryHaney2,SOTTheoryFreimuth,SOTTheoryFreimuth2,SOTTheoryFreimuth3} and drift-diffusion calculations done to analyze experiments.

% ------------------------------------------------------------------------------------------------------------------------------------------------------------------------------------------------------------------
% Conclusions
% ------------------------------------------------------------------------------------------------------------------------------------------------------------------------------------------------------------------

To conclude, we present a theory of spin transport at interfaces with spin-orbit coupling.  The theory describes spin/charge transport in terms of resistive elements, which ultimately describe measurable consequences of interfacial spin-orbit scattering.  In particular, the proposed conductivity and torkivity tensors model the phenomenology of in-plane electric fields in the presence of interfacial spin-orbit coupling, which was previously inaccessible to the drift-diffusion equations.  We calculate all parameters in a simple model, but also provide general expressions in the case of realistic electronic structure.  We found that elements of the conductivity and torkivity tensors are more important than the modifications of other transport parameters (such as the mixing conductance) in many experimentally-relevant phenomena, such as spin orbit torque, spin pumping, the Rashba-Edelstein effect, and the spin Hall magnetoresistance.

% ------------------------------------------------------------------------------------------------------------------------------------------------------------------------------------------------------------------
% Acknowledgements
% ------------------------------------------------------------------------------------------------------------------------------------------------------------------------------------------------------------------

\begin{acknowledgments}
The authors thank Kyoung-Whan Kim, Paul Haney, Guru Khalsa, Kyung-Jin Lee, and Hyun-Woo Lee for useful conversations and Robert McMichael and Thomas Silva for critical readings of the manuscript.  VA acknowledges support under the Cooperative Research Agreement between the University of Maryland and the National Institute of Standards and Technology, Center for Nanoscale Science and Technology, Grant No. 70NANB10H193, through the University of Maryland.
\end{acknowledgments}

% ------------------------------------------------------------------------------------------------------------------------------------------------------------------------------------------------------------------
% Appendix
% ------------------------------------------------------------------------------------------------------------------------------------------------------------------------------------------------------------------

\appendix

% ------------------------------------------------------------------------------------------------------------------------------------------------------------------------------------------------------------------
\section{Derivation of the conductance and conductivity tensors}
\label{ap:SimplifiedForm}
% ------------------------------------------------------------------------------------------------------------------------------------------------------------------------------------------------------------------
%
%
To derive the conductance and conductivity tensors we must approximate the distribution function $\B_{\alpha} (\threevec{k})$ at the interface.  The distribution function gives the momentum-dependent occupancy of carriers described by the spin/charge index $\alpha$.  Just out of equilibrium, it is perturbed by the linearized non-equilibrium distribution function $\LB_{\alpha}(\threevec{k})$, as seen in \eqref{eq:BoltDist}.  In the following we complete the derivation begun in Sec.~\ref{BPDerivation}.  %Carriers belonging to the populations described by $\alpha \in [d,f,\ell,\ch,\uparrow,\downarrow]$ have well-defined Fermi surfaces and velocities; thus we use this spin/charge basis to derive all boundary parameters.  The tensors derived in this section may be represented in other spin/charge bases via straightforward linear transformations.

We write the portion of $\LB_{\alpha}(\kp)$ incident on the interface as done in \eqref{eq:IncDist}.  The first term on the right hand side of \eqref{eq:IncDist} captures the spin and charge currents incident on the interface, while the second term gives an anisotropic contribution caused by an external electric field.  As discussed in Sec.~\ref{BPDerivation}, the simplest approximation for $\LB_{\alpha}(\kp)$ is to use the particular solution of the Boltzmann equation in the relaxation time approximation, given by \eqref{eq:apfSimple}.
The momentum relaxation times that we use account for differing majority ($\tau^\uparrow$) and minority ($\tau^\downarrow$) relaxation times in the ferromagnet, and are renormalized by bulk spin-flip scattering in the non-magnet:
\begin{align}
\label{eq:apTransTimes}
(\tau)^{-1} 	&= (\tau_{\mf})^{-1} + (\tau_{\sfp})^{-1}.
\end{align}
We may better approximate \eqref{eq:apfSimple} by forcing the distribution function to obey outer boundary conditions as well.  In the companion paper we present a more sophisticated approximation for \eqref{eq:apfSimple} that accomplishes this by using solutions to the homogeneous Boltzmann equation.

The outgoing distribution, given by \eqref{eq:OutDist}, is specified by the incoming distribution and the scattering coefficients $S_{\alpha \beta}$.  The scattering coefficients are given by \eqref{eq:SIntUnitary} and \eqref{eq:SIntPrime}.  Here we compute non-equilibrium accumulations analogously to the currents defined by \eqref{eq:DefCurrent},
\begin{align}
\mu_\alpha &= - \frac{1}{e} \frac{1}{A_{FS_\alpha}} \int_{\FS_\alpha} d^2k \LB_{\alpha}(\kvec), 											\label{eq:apDefAccum}		
\end{align}
where $\mu_\alpha$ denotes the accumulation.  Furthermore, $A_{FS_\alpha}$ gives the Fermi surface area while $v_{F_\alpha}$ gives the Fermi velocity.  The quantities just defined apply to the population with spin/charge index $\alpha$.  Likewise, all integrals are evaluated over the Fermi surface that corresponds to the spin/charge index $\alpha$.  Note that we express the accumulations in units of voltage and the current densities in units of number current density.  Using \eqrefs{eq:OutDist}{eq:SIntUnitary} we may rewrite these expressions as integrals over the maximal two-dimensional Brillouin zone common (2DBZ) to all carriers
\begin{align}
\mu_\alpha &=  -\frac{c_\mu}{e} \sum_\beta \int_\TDBZ d\kp \frac{1}{v_{z\alpha}(\kp)} \big{(} \delta_{\alpha\beta} + S_{\alpha\beta} \big{)} \LB^{\i}_{\beta}(\kp)						\label{eq:apDefAccum2DBZ}		\\
j_{i\alpha} &= -\frac{c_j}{e} \sum_\beta \int_\TDBZ d\kp \frac{v_{i\alpha}(\kp)}{v_{z\alpha}(\kp)} \big{(} \delta_{\alpha\beta} \theta_{iz} + S_{\alpha\beta} \big{)} \LB^{\i}_{\beta}(\kp),	\label{eq:apDefCurrent2DBZ}		
\end{align}
where
\begin{align}
c_\mu &\equiv \frac{v_{F\alpha}}{A_{FS_\alpha}},		
\quad \quad
c_j \equiv -\frac{e}{\hbar (2\pi)^3}.		
\end{align}
Note that the velocities correspond to outgoing carriers.  The factor $\theta_{iz} \equiv (1 - 2 \delta_{iz})$ accounts for the fact that incoming and outgoing currents have the opposite sign for $i = z$ but the same sign for $i \in [x,y]$.  By integrating over the maximal two-dimensional Brillouin zone we encounter evanescent states, since $\kp$ vectors not corresponding to real Fermi surfaces have imaginary $k_z$ values.  Here we neglect the contributions to the currents and accumulations due to evanescent states.  Such contributions vanish very close to the interface.

We must now express the accumulations and currents in terms of the incoming polarizations and the in-plane electric field.  Plugging \eqrefs{eq:IncDist}{eq:OutDist} into \eqrefs{eq:apDefAccum2DBZ}{eq:apDefCurrent2DBZ}, we obtain the following
\begin{align}
\mu_{\alpha} &= A_{\alpha \beta} q_\beta + a_{\alpha} \EF 			\label{eq:apMuQ}	\\
j_{i\alpha} &= B_{i\alpha \beta} q_\beta + b_{i\alpha} \EF			\label{eq:apJQ}
\end{align}
where the tensors that contract with the incident spin/charge polarization are given by
\begin{align}
A_{\alpha\beta} &= c_\mu \int_{\TDBZ} d\kp \frac{1}{v_{z\alpha}} \big{(} \delta_{\alpha\beta} + S_{\alpha\beta} \big{)}										\label{eq:apATensor}		\\
B_{i\alpha\beta} &= c_j \int_{\TDBZ} d\kp \frac{v_{i\alpha}}{v_{z\alpha}} \big{(} \delta_{\alpha\beta} \theta_{iz} + S_{\alpha\beta} \big{)} 						\label{eq:apBTensor}	
\end{align}
while the tensors that multiply the in-plane electric field become
\begin{align}
a_{\alpha} &= c_\mu \sum_\beta \int_{\TDBZ} d\kp \frac{1}{v_{z\alpha}} \big{(} \delta_{\alpha\beta} + S_{\alpha\beta} \big{)} \LBE_{\beta}							\label{eq:apaTensor}		\\
b_{i\alpha} &= c_j \sum_\beta \int_{\TDBZ} d\kp \frac{v_{i\alpha}}{v_{z\alpha}} \big{(} \delta_{\alpha\beta} \theta_{iz} + S_{\alpha\beta} \big{)} \LBE_{\beta}			\label{eq:apbTensor}	
\end{align}
In the same spirit as magnetoelectronic circuit theory, these tensors represent moments of the scattering coefficients weighted by velocities.

To determine exactly how the currents depend on the accumulations, we solve for $j_{i\alpha}$ in terms of $\mu_\alpha$.   Doing so yields the following conductance and conductivity tensors
\begin{align}
%\label{eq:apCurrentGeneral}
G_{i\alpha\beta} &= B_{i\alpha\gamma} [A^{-1}]_{\gamma\beta}	\nonumber\\
\sigma_{i\alpha} &= b_{i\alpha} - G_{i\alpha\beta} a_{\beta}.	\nonumber
\end{align}
To further simplify these expressions, we follow the example of magnetoelectronic circuit theory \cite{MCTBrataas, MCTBrataas2} and assume that the incoming spin-currents behave as if they originate from spin-dependent reservoirs.  This implies that the incoming spin polarization $q_\alpha$ equals the quasichemical potential $\mu_\alpha$ at the interface.  For this to be true, we must find that $A_{\alpha\beta} \propto \delta_{\alpha\beta}$ and $a_{\alpha} = 0$ by inspection of \eqref{eq:apMuQ}.  These relationships hold if one evaluates \eqrefs{eq:apATensor}{eq:apaTensor} over the incoming portion of the Fermi surface only.  We find that the contributions from the outgoing portion of the Fermi surface cancel to a good approximation, which suggests that:
\begin{align}
\label{eq:apCurrentSimple}
G_{i\alpha\beta} &= B_{i\alpha\beta}			\\
\sigma_{i\alpha} &= b_{i\alpha}				.		
\end{align}
The above equations give simpler expressions for the conductance and conductivity tensors in terms of interfacial scattering coefficients.

% ------------------------------------------------------------------------------------------------------------------------------------------------------------------------------------------------------------------
\section{Derivation of the torkance and torkivity tensors}
\label{ap:DerivationTorques}
% ------------------------------------------------------------------------------------------------------------------------------------------------------------------------------------------------------------------
%
%
To describe the spin torque at $z = 0$, we must compute the ensemble average of spin density $\langle s_\sigma \rangle$ using \eqref{eq:DefTorque}.  The resulting torque is given by \eqref{eq:DefTorque2}. To describe the spin torque at $z = 0^+$, we must calculate the transverse spin current in the ferromagnet using \eqrefs{eq:DefTorqueFM}{eq:TransverseVelocityFM}.  We then express the spin torque in terms of the incoming polarizations and the in-plane electric field by plugging \eqrefs{eq:IncDist}{eq:OutDist} into \eqrefsss{eq:DefTorque}{eq:DefTorque2}{eq:DefTorqueFM}.  In doing so we obtain 
\begin{align}
\tau_{\sigma} &= C_{\sigma \beta} q_\beta + c_{\sigma} \EF, 		\label{eq:TauapQ}
\end{align}
where
\begin{align}
C_{\sigma \beta}	&=		C^\FM_{\sigma \beta} + C^\magnetization_{\sigma \beta} 				\label{eq:apSplitC}		\\
c_{\sigma} 		&=		c^\FM_{\sigma} + c^\magnetization_{\sigma} 							\label{eq:apSplitC}		
\end{align}
describes the separation of the spin torque into its bulk ferromagnet and interface contributions.  The tensors that contract with the incident spin/charge polarization are given by
\begin{align}
C^\FM_{\sigma\beta} = c_j &\int_{\TDBZ} d\kp \frac{\avg{v}_{z}}{v_{z\beta}} \bar{S}_{\sigma\beta},									\label{eq:apCTensorFM}	\\
C^\magnetization_{\sigma\beta} = - \frac{J_\text{ex}}{\hbar} c_j \sum_{\sigma'} &\int_{\TDBZ} d\kp \frac{1}{v_{z\beta}} \epsilon_{\sigma\sigma'} T_{\sigma'\beta},						\label{eq:apCTensorInt}		
\end{align}
while the tensors that multiply the in-plane electric field become
\begin{align}
c^\FM_{\sigma} = c_j \sum_\beta &\int_{\TDBZ} d\kp \frac{\avg{v}_{z}}{v_{z\beta}} \bar{S}_{\sigma\beta} \LBE_{\beta}	.						\label{eq:apcTensorFM}		\\
c^\magnetization_{\sigma} = - \frac{J_\text{ex}}{\hbar} c_j \sum_\beta &\int_{\TDBZ} d\kp \frac{1}{v_{z\beta}} \epsilon_{\sigma\sigma'} T_{\sigma'\beta} \LBE_{\beta}				\label{eq:apcTensorInt}		
\end{align}
where the velocity $\avg{v}_{z}(\kp)$ corresponds to the outgoing portion of the Fermi surface in the ferromagnet.

As we did for the currents, we solve for $\tau_\sigma$ in terms of $\mu_\alpha$.   Doing so yields the following torkance and torkivity tensors
\begin{align}
%\label{eq:apTorqueGeneral}
\Gamma_{\sigma\beta} &= C_{\sigma\gamma} [A^{-1}]_{\gamma\beta}	\nonumber\\
\gamma_{\sigma} &= c_{\sigma} - \Gamma_{\sigma\beta} a_{\beta}. \nonumber
\end{align}
The torkance tensor describes the contribution to the total spin torque that arises from the accumulations at the interface.  The torkivity tensor describes the subsequent contribution from interfacial spin-orbit scattering when driven by an in-plane electric field.  Following the arguments made for \eqref{eq:apCurrentSimple}:
\begin{align}
\label{eq:apTorqueSimple}
\Gamma_{\sigma\beta} &= C_{\sigma\beta}		\\
\gamma_{\sigma} &= c_{\sigma}.
\end{align}
As seen in the companion paper, this approximation produces good agreement with the interfacial charge currents, spin currents, and spin torques computed via the Boltzmann equation.

% ------------------------------------------------------------------------------------------------------------------------------------------------------------------------------------------------------------------
\section{Boundary Parameters for Realistic Interfaces}
\label{ap:GeneralForm}
% ------------------------------------------------------------------------------------------------------------------------------------------------------------------------------------------------------------------
%
%
To generalize the expressions from the previous section to include electronic structure, we must consider the non-equilibrium distribution function for all bands relevant to transport:
\begin{align}
\B_{m\alpha} (\threevec{k}) = \B^{\eq}_{m\alpha}(\varepsilon_{m\alpha\threevec{k}}) + \pdf{\B^{\eq}_{m\alpha}}{\varepsilon_{m\alpha\threevec{k}}} \LB_{m\alpha}(\threevec{k}).
\label{eq:apBoltDist}
\end{align}
Here $m$ describes the spin-independent band number and $\alpha$ denotes the spin/charge index.  If the case of a non-magnet, for each spin-independent band there are two degenerate states.  Linear combinations of these states can produce phase coherent spin states that point in any direction.  Thus, for the non-magnet, the spin/charge index should span $\alpha \in [d,f,\ell,\ch]$, where the $\ell$ direction is aligned with the magnetization in the neighboring ferromagnet for convenience.  In the ferromagnet all bands are non-degenerate, so each state possesses a different phase velocity.  As a result, linear combinations of these states have spin expectation values that oscillate over position, complicating the description presented above.  There is no natural pairing of non-degenerate spin states.  However, if states are quantized along a particular axis, the spin accumulations and spin currents with polarization along that axis are well-defined regardless of the choice of pairing.  Thus for each spin-independent band in the ferromagnet, the spin/charge index spans the states describing majority and minority carriers, i.e. $\alpha \in [\uparrow, \downarrow]$.

We generalize the approximate distribution function $\LBE_{m\alpha}(\kp)$ caused by an external electric field to allow for a band dependence.  We do so because the velocities now depend on band number and the scattering times may as well.  However, we assume that the incoming polarization $q_\alpha$ does not depend on band number; thus we treat incident currents as if they originate from spin-dependent (but not band-dependent) reservoirs.  The momentum relaxation times for each spin-independent band in the non-magnet are renormalized using \eqref{eq:apTransTimes}.

To account for coherence between bands, we begin with a more general expression for the ensemble average of the outgoing current:
\begin{align}
\langle \langle j^\o_{i\alpha} \rangle \rangle	 = \frac{1}{\hbar} \sum_{mnn'\beta}	\int_\TDBZ		&d\kp \frac{ \LB^{\i}_{m\beta}}{|v_{mz\beta}|}											\nonumber	\\
																					&\times \mathrm{tr} \big{[} (s_{n'm})^\dagger J^\o_{n'n,i\alpha} s_{nm} \sigma_\beta \big{]}		.
\nonumber \\ \nonumber \\[-0.2cm]
\label{eq:apOutCurrentGen}
\end{align}
Here $s$ stands for reflection or transmission, depending on what region(s) incoming and outgoing carriers are from.  The indices $m$ and $\beta$ correspond to incoming carriers, while $n$, $n'$, and $\alpha$ describe the outgoing carriers.  The current operator $J^\o_{n'n,i\alpha}$ is given by
\begin{align}
J^\o_{n'n,i\alpha} = \frac{i\hbar}{2m} \int_{\TDPC} d\rp (\Psi_{n'\kvec})^\dagger (\Lpartial{i} \sigma_\alpha - \sigma_\alpha \Rpartial{i} ) \Psi_{n\kvec}
\nonumber \\
\label{eq:apOutCurrentOpGen}
\end{align}
where the integral runs over a two-dimensional slice of the primitive cell (aligned parallel to the interface).  The $2 \times 2$ matrix $\Psi_{n\kvec}$ is defined for outgoing modes in the ferromagnet as
\begin{align}
\label{eq:apPsiOpGen}
\Psi_{n\kvec} = e^{i \kp \cdot \rp}
\begin{pmatrix}
u^\uparrow_{n\kvec}(\rvec) e^{i k^\uparrow_{nz} z}		&	0														\\
0													&	u^\downarrow_{n\kvec}(\rvec) e^{i k^\downarrow_{nz} z}				
\end{pmatrix},
\end{align}
where $u^{\uparrow/\downarrow}_{n\kvec}(\rvec)$ and $k^{\uparrow/\downarrow}_{nz}$ denote the Bloch wavefunction and out-of-plane wavevector for majority/minority carriers.  Both are defined at $\kp$ on the Fermi surface corresponding to band $n$.  For outgoing modes in the non-magnet, $\Psi_{n\kvec}$ simplifies to:
\begin{align}
\label{eq:apPsiOpGen}
\Psi_{n\kvec} = e^{i \kp \cdot \rp} e^{i k_{nz} z} u_{n\kvec}(\rvec) I_{2 \times 2}		.
\end{align}
The incoming current is defined as follows
\begin{align}
\langle \langle j^\i_{i\alpha} \rangle \rangle	 = \frac{1}{\hbar} \sum_{m\beta}	\int_\TDBZ	&d\kp \frac{ \LB^{\i}_{m\beta}}{|v_{mz\beta}|}	\mathrm{tr} \big{[} J^\i_{m,i\alpha} \sigma_\beta \big{]}
\nonumber \\
\label{eq:apInCurrentGen}
\end{align}
where
\begin{align}
J^\i_{m,i\alpha} = \theta_{iz} \frac{i\hbar}{2m} \int_{\TDPC} d\rp (\Psi_{m\kvec})^\dagger (\Lpartial{i} \sigma_\alpha - \sigma_\alpha \Rpartial{i} ) \Psi_{m\kvec}
\nonumber \\[0.1cm]
\label{eq:apInCurrentOpGen}
\end{align}
gives the current operator for the incoming current.  The total current is then
\begin{align}
\langle \langle j_{i\alpha} \rangle \rangle	&=	\langle \langle j^\i_{i\alpha} \rangle \rangle		+		\langle \langle j^\o_{i\alpha} \rangle \rangle				\nonumber \\
									&= \frac{1}{\hbar} \sum_{mnn'\beta}	\int_\TDBZ	d\kp \frac{ \LB^{\i}_{m\beta} }{|v_{mz\beta}|}			\nonumber \\
									&\times \mathrm{tr} \Big{[} \big{(} \theta_{iz} J^\i_{m,i\alpha} + (s_{n'm})^\dagger J^\o_{n'n,i\alpha} s_{nm} \big{)} \sigma_\beta \Big{]}
\nonumber \\ \nonumber \\[-0.2cm]
\label{eq:apTotalCurrentGen}
\end{align}
where the choice of scattering matrix depends on the incoming spin/charge index $\beta$ and outgoing spin/charge index $\alpha$ as follows:
\begin{align}
s_{nm} =
\begin{cases} 
r_{nm}					~~\quad \quad 		\alpha, \beta \in [d,f,\ell,\ch]										\\[0.15cm]
t_{nm}					~~\quad \quad \:\!	\alpha \in [d,f,\ell,\ch], ~~ \beta \in [\uparrow,\downarrow]			\\[0.15cm]
t^*_{nm}				~~\quad \quad \:\!	\alpha \in [\uparrow,\downarrow], ~~ \beta \in [d,f,\ell,\ch]			\\[0.15cm]
r^*_{nm}				~~\quad \quad 		\alpha, \beta \in [\uparrow,\downarrow]							
\end{cases} \nonumber \\
\label{eq:apSIntGen}
\end{align}
By plugging in the generalizations of \eqrefs{eq:IncDist}{eq:apfSimple} into \eqref{eq:apTotalCurrentGen}, we find that \eqref{eq:apBTensor} generalizes to the following
\begin{align}
B_{i\alpha\beta}			&= -\frac{e}{\hbar} \sum_{mnn'}		\int_\TDBZ	d\kp \frac{ 1 }{|v_{mz\beta}|}													\nonumber \\
						&\times \mathrm{tr} \Big{[} \big{(} \theta_{iz} J^\i_{m,i\alpha} + (s_{n'm})^\dagger J^\o_{n'n,i\alpha} s_{nm} \big{)} \sigma_\beta \Big{]}		,
\nonumber \\ \nonumber \\[-0.25cm]			
\label{eq:apBTensorGen}	
\end{align}
while \eqref{eq:apbTensor} now becomes:
\begin{align}
b_{i\alpha}				&= -\frac{e}{\hbar} \sum_{mnn'\beta}		\int_\TDBZ	d\kp \frac{ \LBE_{m\beta} }{|v_{mz\beta}|}													\nonumber \\
						&\times \mathrm{tr} \Big{[} \big{(} \theta_{iz} J^\i_{m,i\alpha} + (s_{n'm})^\dagger J^\o_{n'n,i\alpha} s_{nm} \big{)} \sigma_\beta \Big{]}		.
\nonumber \\ \nonumber \\[-0.25cm]			
\label{eq:apbTensorGen}	
\end{align}
Assuming as before that the incoming spin-currents behave as if they originate from spin-dependent reservoirs ($\mu_\alpha \approx q_\alpha$), we have:
\begin{align}
G_{i\alpha\beta} &= B_{i\alpha\beta}			\\
\sigma_{i\alpha} &= b_{i\alpha}				.		
\end{align}
Thus, \eqrefs{eq:apBTensorGen}{eq:apbTensorGen} generalize the conductance and conductivity tensors respectively to include non-trivial electronic structure.  

The transverse spin current that develops in the ferromagnet at $z = 0^+$ may be obtained by using similar expressions.  The tensor $C^\FM_{\sigma\beta}$, originally given by \eqref{eq:apCTensorFM}, now becomes
\begin{align}
C^\FM_{\sigma\beta}		&= -\frac{e}{\hbar} \sum_{mnn'}		\int_\TDBZ	d\kp \frac{ 1 }{|v_{mz\beta}|}												\nonumber \\[0.15cm]
						&\times
\begin{cases} 
\mathrm{tr} \big{[} (t^*_{n'm})^\dagger J^\o_{n'n,z\sigma} t^*_{nm} \sigma_\beta \big{]}			\quad \quad \,			\beta \in [d,f,\ell,\ch]				\\[0.3cm]
\mathrm{tr} \big{[} (r^*_{n'm})^\dagger J^\o_{n'n,z\sigma} r^*_{nm} \sigma_\beta \big{]}			\quad \quad				\beta \in [\uparrow,\downarrow]		.	
\end{cases}
\nonumber \\ \nonumber \\[-0.25cm]			
\label{eq:apCFMTensorGen}	
\end{align}
Likewise, the tensor $c^\FM_{\sigma}$, first described by \eqref{eq:apcTensorFM}, generalizes to the following:
\begin{align}
c^\FM_{\sigma}			&= -\frac{e}{\hbar} \sum_{mnn'\beta}	\int_\TDBZ	d\kp \frac{ \LBE_{m\beta} }{|v_{mz\beta}|}								\nonumber \\[0.15cm]
						&\times
\begin{cases} 
\mathrm{tr} \big{[} (t^*_{n'm})^\dagger J^\o_{n'n,z\sigma} t^*_{nm} \sigma_\beta \big{]}			\quad \quad \,			\beta \in [d,f,\ell,\ch]				\\[0.3cm]
\mathrm{tr} \big{[} (r^*_{n'm})^\dagger J^\o_{n'n,z\sigma} r^*_{nm} \sigma_\beta \big{]}			\quad \quad				\beta \in [\uparrow,\downarrow]		.	
\end{cases}
\nonumber \\ \nonumber \\[-0.25cm]			
\label{eq:apcFMTensorGen}	
\end{align}
Evaluating the trace in \eqref{eq:apcFMTensorGen} gives the ensemble average of velocity for the transverse spin states in the ferromagnet.  Here we do not assume that the velocity of these states equals the average velocity of majority and minority carriers.  However, for the simple model discussed in the previous section, one can show that the current operator $J^\o_{n'n,z\sigma}$ simplifies to the following:
\begin{align}
J^\o_{n'n,z\sigma}	\rightarrow	J^\o_{z\sigma}	\propto	\frac{1}{2} \big{(} v_{z\uparrow} + v_{z\downarrow} \big{)} \sigma_\sigma		
\end{align}
In this scenario, \eqrefs{eq:apCFMTensorGen}{eq:apcFMTensorGen} reduce to \eqrefs{eq:apCTensorFM}{eq:apcTensorFM} as expected.  This justifies the use of the average velocity to describe transverse spin states in the simple model.  For $\mu_\alpha \approx q_\alpha$ we have:
\begin{align}
\Gamma^\FM_{\sigma\beta} &= C^\FM_{\sigma\beta}			\\
\gamma^\FM_{\sigma} &= c^\FM_{\sigma}					.		
\end{align}
Thus we have generalized the torkance and torkivity tensors that describe bulk ferromagnet torques for non-trivial electronic structures.

For realistic systems, the interface should be modeled over a few atomic layers so that an exchange potential and spin-orbit coupling may simultaneously exist.  If these atomic layers make up the scattering region used to obtain the scattering coefficients, then the expressions presented here describe the currents on either side of the interface as intended.  However, in order to describe the interfacial torque, the tensors $C^\magnetization_{\sigma\beta}$ and $c^\magnetization_{\sigma}$ must be written as sums of the layer-resolved torques within the interfacial scattering region.  We save the generalization of \eqrefs{eq:apCTensorInt}{eq:apcTensorInt} for future work, since in this paper we treat the interface as a plane rather than a region of finite thickness.

\section{Boundary parameters relevant to bilayer spin-orbit torques}

% ------------------------------------------------------------------------------------------------------------------------------------------------------------------------------
% Table: Relevant Boundary Parameters for DDE Solution
% ------------------------------------------------------------------------------------------------------------------------------------------------------------------------------
\begin{table}
	\begin{tabular}{ l | l  }
	\textbf{Parameter}		&	\textbf{Value}	\\ 
	\hline
	 \multicolumn{2}{l}{Effective mixing conductance} \\
	$\ReGem$			&	$G_{z d d}$ ~or~ $G_{z f f}$														\\
	$\ImGem$				&	$G_{z d f}$ ~or~ $G_{z f d}$														\\[0.1cm] 
	\hline
	\multicolumn{2}{l}{Spin current due to interfacial spin-orbit scattering} \\
	$j^\RE_{d}(0^-)$		&	$\sigma_{z d} \EF$								\\
	$j^\RE_{f}(0^-)$			&	$\sigma_{z f} \EF$								\\[0.1cm] 
	\hline
	\multicolumn{2}{l}{Spin torque on the lattice at the interface} \\
	$\tau^\lattice_{d}$		&	$\big{(} \sigma_{z d} - \gamma_{d} \big{)} \EF$									\\
	$\tau^\lattice_{f}$		&	$\big{(} \sigma_{z f} - \gamma_{f} \big{)} \EF$										\\
 	\end{tabular}
	\caption{
	Table of phenomenological parameters relevant to the drift-diffusion model of spin-orbit torque developed in the companion paper, chosen by the numerical study performed in Sec.~$\ref{BPAnalysis}$.  All other boundary parameters are discarded in that model.  As can be seen in section IIIA of the companion paper, the first four parameters govern the total spin torque thickness dependence, while the last two parameters describe the spin torque's zero-thickness intercept.  Note that here all boundary parameters obey the sign convention that positive currents flow towards from the ferromagnet.
	}
	\label{tb:BP}
\end{table}
% ------------------------------------------------------------------------------------------------------------------------------------------------------------------------------
% End Table
% ------------------------------------------------------------------------------------------------------------------------------------------------------------------------------

In Sec.~\ref{BPAnalysis}, we numerically analyze each boundary parameter for an interfacial scattering potential that includes the exchange interaction and spin-orbit coupling.  We find that many parameters differ by several orders of magnitude.  In the companion paper, we use this information to derive an analytical drift-diffusion model of spin-orbit torques in heavy metal/ferromagnet bilayers.  In the following we discuss the minimal set of parameters crucial to that solution.

Table \ref{tb:BP} includes six parameters important to the interface of heavy metal/ferromagnet bilayers.  Along with the spin diffusion length ($\lsf$), the bulk conductivity ($\sigma^\NM_\bulk$), and the spin Hall current density ($j^\SHE_{d}$) in the non-magnet, they describe all of the phenomenological parameters used by the analytical drift-diffusion model in the companion paper.  The first two parameters are the real and imaginary parts of the spin mixing conductance.  The generalized version of these parameters may be extracted from the conductance tensor $G_{z\alpha\beta}$.  Numerical studies show that these parameters depend weakly on magnetization direction.  In the companion paper, the ungeneralized spin mixing conductance is used.  The parameters $j^\RE_{d}(0^-)$ and $j^\RE_{f}(0^-)$ denote the interfacial spin currents just within the non-magnet that arise due to in-plane electric fields and spin-orbit scattering.  In analogy to the bulk spin Hall current, these parameters act as sources of spin current for the drift-diffusion equations.  Thus, in the absence of $j^\RE_{d}(0^-)$, $j^\RE_{f}(0^-)$, and $j^\SHE_{d}$, all bulk currents and accumulations vanish.  In addition to the spin mixing conductance, these parameters determine the non-magnet thickness-dependence of spin-orbit torques.  The final two parameters give the approximate loss of angular momentum to the interface.  They equal the damping-like and field-like spin-orbit torques in the limit of vanishing non-magnet thickness.  They are derived by subtracting the interfacial torque from the loss in out-of-plane spin current density across the interface.  Our numerical analysis suggests that spin and charge accumulations cause negligible differences in these two quantities.  Thus, we assume that $\tau^\lattice_{d}$ and $\tau^\lattice_{f}$ stem primarily from spin-orbit scattering at the interface.  The treatment of the lattice torque presented in the companion paper begins from this assumption.

The model introduced in the companion paper generalizes the drift-diffusion model used in Ref.~\cite{SOTTheoryHaney} to include interfacial spin-orbit effects.  Only two additional phenomenological parameters ($j^\RE_{d}(0^-)$ and $j^\RE_{f}(0^-)$) are required to capture the non-magnet thickness dependence, while an additional two parameters ($\tau^\lattice_{d}$ and $\tau^\lattice_{f}$) describe the corresponding zero-thickness intercept.  Table \ref{tb:BP} provides formulas for these phenomenological parameters in terms of the boundary parameters contained within \eqrefs{eq:BC1}{eq:BC2}.  We note that in magnetoelectronic circuit theory, the conductance parameters are given by sums of interfacial scattering coefficients over the available scattering states.  All of the boundary parameters introduced here possess a similar form, as discussed in appendices \ref{ap:SimplifiedForm}, \ref{ap:DerivationTorques}, and \ref{ap:GeneralForm}.  

% ------------------------------------------------------------------------------------------------------------------------------------------------------------------------------------------------------------------
% Bibliography
% ------------------------------------------------------------------------------------------------------------------------------------------------------------------------------------------------------------------

\bibliography{apssamp}% Produces the bibliography via BibTeX.

% ------------------------------------------------------------------------------------------------------------------------------------------------------------------------------------------------------------------
% End of document
% ------------------------------------------------------------------------------------------------------------------------------------------------------------------------------------------------------------------

\end{document}